\documentclass[fractalfract,review,accept,pdftex,oneauthor]{Definitions/mdpi} 

\firstpage{1} 
\pubvolume{9}
\issuenum{1}
\articlenumber{579}
\pubyear{2025}
\copyrightyear{2025}
\externaleditor{Firstname Lastname} 
\datereceived{9 May 2025} 
\daterevised{17 June 2025} 
\dateaccepted{22 August 2025} 
\datepublished{31 August 2025 } 
\hreflink{https://doi.org/10.3390/
  \linebreak
  fractalfract9090579} 

\usepackage{soul}
\usepackage{graphicx,rotating,amsmath,amsfonts,amssymb}

\usepackage{natbib}
\usepackage[utf8]{inputenc}
\newcommand{\Gpc}{$h^{-1}$\thinspace Gpc} 
\newcommand{\Mpc}{$h^{-1}$\thinspace Mpc}

\newcommand{\be}{\begin{equation}}
\newcommand{\ee}{\end{equation}}
\newcommand{\dd}[1]{{\rm d}#1\,}

\newcommand{\ba}{\begin{array}} 
\newcommand{\ea}{\end{array}}

\def\apjl{ApJL}
\def\apjs{ApJS}

\def\aap{A\&A}

\def\prl{Phys.~Rev.~Lett.}%

\Title{Fractal Properties of the Cosmic Web}

\TitleCitation{Fractal Properties of the Cosmic Web}

\Author{{Jaan Einasto} 
} 

\AuthorNames{Jaan Einasto}

\AuthorCitation{Einasto, J.}

\address[1]{Tartu Observatory, University of Tartu, Observatooriumi 1, 61602 T\~oravere, Estonia; 
{jaan.einasto@ut.ee} 
}

\abstract{The cosmic web is one of the most complex systems in nature,
  consisting of galaxies and clusters of galaxies joined by filaments
  and walls, leaving large empty regions called cosmic voids. The
  most common method of describing the web is a correlation function and
  its derivative, the fractal function. In this paper, I provide a
  review of the fractal properties of the cosmic web from the
  observational point of view within the Newtonian concordance
  $\Lambda$CDM Universe framework.  I give a brief history of fractal
  studies of the Universe. I then describe the derivation of the
  fractal function from angular and spatial distributions of galaxies
  and their relations.  Correlation functions are not sensitive to the
  shape of the galaxy distribution. To improve our quantitative
  understanding of properties of the web, statistics must be used
  which are sensitive to the pattern of the web.  }

\begin{document}

\section{Introduction \label{intro}}

Long ago scientists noticed that many natural processes are
self-similar over a large range of scales. Well-known examples are
coastlines and mountain regions.  The~self-similarity of natural
processes was discussed by Benoit \citet{Mandelbrot:1977aa}, who
suggested the term {\em fractal} for this phenomenon.  A~similar
phenomenon was observed in the distribution of galaxies, which are
hierarchically clustered.  This was noticed  by
\citet{Charlier:1922aa} and studied in more detail by
\citet{Carpenter:1938aa}, \citet{Kiang:1967}
\citet{Wertz:1970aa,Wertz:1971aa}, \mbox{\citet{Haggerty:1972aa}} and
\citet{de-Vaucouleurs:1970, de-Vaucouleurs:1971aa}.  Based on these 
observations a branch in physical cosmology, named fractal cosmology,
was formed in the 1980s \citep{Pietronero:1987aa}.  An~important
issue in the fractal cosmology is the fractal dimension and its
dependence on the scale. The~fractal dimension of a homogeneous
spatial object is three, of~a surface is two, and~of a line is one. Actual
objects can have non-integer values of the fractal~dimension.

Studies of the fractal properties of the cosmic web are conducted using either the
Newtonian framework or the relativistic approach. In~the Newtonian framework,
researchers use direct observational data and N-body simulations to
describe large-scale structures, assuming gravity to be the dominant 
force, without~accounting for relativistic effects. This approach
treats cosmic structures within the concordant Lambda Cold Dark Matter
($\Lambda$CDM) 
universe model, relying on classical mechanics to model clustering patterns.
The history of the formation of this model is given, among others, in
books by \citet{Peebles:2020aa, Peebles:2022aa} and \citet{Einasto:2024aa}.

In contrast, the~relativistic approach incorporates general
relativity, taking into account the expansion of the Universe and
relativistic corrections to gravitational interactions. This
perspective provides a more accurate description of cosmic evolution,
particularly on a large scale, where relativistic effects influence
the formation of the distribution of structures. Studies in this field
examine how fractal-like properties emerge within the relativistic
framework.  This often involves the use of tensor-based models and
relativistic perturbation theory.  Relativistic approaches include,
among other topics, theories of inflation \citep{Starobinsky:1980ys,
  Guth:1981}, chaotic inflation (\citet{Linde:1983yq, Linde:1994aa,
  Linde:1997aa, Linde:2005aa}, and \mbox{\citet{Nambu:1989aa})}, and quantum
gravity (\citet{Ambjorn:2005aa} and \citet{Calcagni:2010aa,
  Calcagni:2010ab}).

In this review, I provide an overview of the fractal properties of the
cosmic web in the Newtonian approximation of the Lambda Cold Dark
Matter ($\Lambda$CDM) Universe.  The~review is based on the Newtonian
approach for two reasons: (i) almost all fractal studies of real and
simulated galaxies are conducted within the framework of the Lambda
Cold Dark Matter universe, described in a new section; (ii) the
relativistic approach is mostly related to the early stages of the
evolution of the Universe, where some constraints of the concordant
Lambda Cold Dark Matter Universe are invalid.  The~relativistic
approach is a new and rapidly evolving field of study.  It is outside
the scope of the present observational~review.

I begin with the description of the Lambda Cold Dark Matter
($\Lambda$CDM) Universe in Section~\ref{LCDM}.  Next, I give a brief
history of fractal studies of the cosmic web in
Section~\ref{history}. Section~\ref{angular} discusses the angular
distribution of galaxies and how this can be described by the angular
correlation function.  Section~\ref{spatial} discusses the statistical
description of the cosmic web by measuring the correlation function
and fractal dimension.  In~Sections~\ref{cf} and \ref{web}, I discuss
the correlation and fractal analysis of the web using spatial data.
Section~\ref{2d3d} is devoted to comparing angular and spatial
distributions of galaxies. Here, I pay special attention to two
aspects of fractal studies: the dependence of fractal characteristics
on the scale from sub-megaparsec to hundreds of megaparsecs and~the
differences between 2D and 3D fractal characteristics. 
Section~\ref{velocities} is devoted to the study of the structure and
evolution of the cosmic web, using combined spatial and velocity data.
Section~\ref{scale} discusses the scale of homogeneity of the cosmic
web.  The~review concludes with a summary and~outlook.

\section{Basics of the Concordant \boldmath{$\Lambda$}CDM~Universe \label{LCDM}}

In Section 2, I  
 describe the concordant  $\Lambda$CDM Universe, the~basic framework
 of fractal studies of the cosmic web.   
The concordant $\Lambda$CDM model of the Universe is based on five
pillars: the Big Bang model of the birth of the Universe, the~Big Bang
nucleosynthesis, data on the cosmic microwave background (CMB) radiation,
data on the web-like distribution of galaxies in the present epoch, and~the
inflation~hypothesis.

The Big Bang model is based on the general relativity  theory by
\citet{Einstein:1916zr} and its extensions,  developed by
\citet{Friedmann:1924ys} and \citet{Lemaitre:1927lh}. An~alternative
model of a Steady-state Universe by \citet{Hoyle:1948} contradicts
many astronomical data and is now rejected. The~physics of the Big
Bang is now well known. There exist variants that suggest that the
Bang that created our Universe was actually only one event in the chaotic
inflation, as~discussed, among others, by \citet{Linde:1983yq,
  Linde:2005aa}.

According to the Big Bang model, the~Universe began in an extremely
hot and dense state. After~a few minutes, the Universe cooled to
temperatures that allowed  light chemical elements---hydrogen, helium
and deuterium---to form. This process is called Big Bang 
nucleosynthesis and~was studied first by \mbox{\citet{Hoyle:1946aa}}
and~more recently by \mbox{\citet{Cyburt:2003aa}}.  All heavier
elements were 
synthesized in stars, as~studied in detail by \mbox{\citet{Burbidge:1957aa}}.
The results of these calculations are in good agreement with the observed
distribution of chemical elements in stars and gas~clouds.

\begin{figure}[H]
  \resizebox{0.95\textwidth}{!}{\includegraphics*{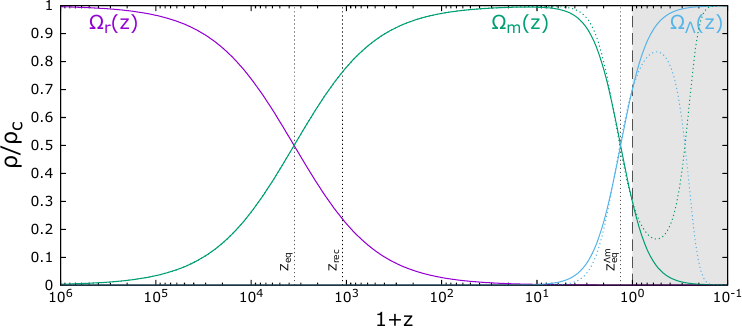}}
  \caption{The   evolution of radiation ($\Omega_{\rm R}(z)$), matter
    ($\Omega_{\rm m}(z)$), and dark energy  ($\Omega_{\Lambda}(z)$) 
    densities shown as a function of redshift $z$
    \citep{Einasto:2025aa}. Reproduced with permission from
    J. Einasto, G. H\"utsi, I. Szapudi, P. Tenjes, Spinning the
    Cosmic Web; published by World Scientific, 2025. 
  }
  \label{fig:LCDM_evol}
\end{figure}

The evolution of densities of various components of the Universe in
units of the critical density is shown in Figure~\ref{fig:LCDM_evol}.
The total density is equal to the critical density with very high
accuracy, since even small deviations from the critical density
increase during the evolution.  The~vertical dashed line corresponds
to the present moment, and the gray shaded region represents the future.  The vertical
dotted lines show epochs of equality of radiation and matter,
$z_{eq}$, recombination, $z_{rec}$, and~equality of dark energy and
matter, $z^{LM}_{eq}$.  Solid colored lines show components of the standard
$\Lambda$CDM, and dotted lines represent a model, where $\Lambda$ is
replaced by decaying dark energy, as~suggested by recent DESI
measurements~\citep{DESI-Collaboration:2024ad}.

The third important epoch in the cosmic history is the recombination of
hydrogen at $z\approx 1000$ at temperatures around 3000$^\circ$~K. The~emission from this epoch is observable as CMB radiation. As~stressed
by \citet{Sunyaev:2009aa}, the~physics at this epoch is very simple
and well understood from laboratory experiments. The CMB radiation angular
power spectrum depends on essential cosmological parameters. Modern
CMB observations with the Planck satellite
\citep{Planck-Collaboration:2020aa} yield for the spatial curvature of
the Universe $\Omega_k = 0.0007\pm 0.0019$.  This means that the Planck
data did not find any deviations from a spatial flat Universe with
$\Omega_k=0$ and $\Omega_{tot}=1$. For~the amount of matter, the Planck
data give the following values: baryon density
$\Omega_b\,h^2=0.02233\pm 0.00015$, cold dark matter (CDM) density
\mbox{$\Omega_ch^2=0.1198\pm 0.0012$}, and~dark energy density
$\Omega_\Lambda =0.6889\pm 0.0056$.  These density estimates are in
good agreement with estimates from Big Bang nucleosynthesis for
baryonic matter, dark matter (DM) in systems of galaxies, as~found
by \citet{Einasto:1974} and \citet{Ostriker:1974},  and~with the 
dark energy density value found from direct measurements of supernovas by
\citet{Perlmutter:1999} and \citet{Riess:1998aa}. Planck and recent
James Webb Space Telescope data yield for the Hubble constant
$H_0= 70.2 \pm 1.4$ km s$^{-1}$Mpc$^{-1}$, and~the age of the Universe
$t_0=13.77\pm 0.12$ Gyr \citep{Komatsu:2011zr},

The inflation hypothesis of the early evolution of the Universe was
suggested independently by \citet{Starobinsky:1980ys} and
\citet{Guth:1981} and~extended by \citet{Linde:1983yq,Linde:2005aa}
to chaotic inflation.  Possible problems of the concordant
$\Lambda$CDM model were analyzed in detail by
\citet{Di-Valentino:2025aa}.

\section{A Short History of the Fractal Studies  of the  Cosmic~Web  \label{history}}

In this section, I give a short history of fractal studies of the
cosmic web.  First, I describe the angular distribution of galaxies and
the discovery of the cosmic web.  Discussion of the fractal
character of the cosmic~web follows.

\subsection{Angular Distribution of~Galaxies \label{angular}}

\begin{figure}[H]

\resizebox{0.60\textwidth}{!}{\includegraphics*{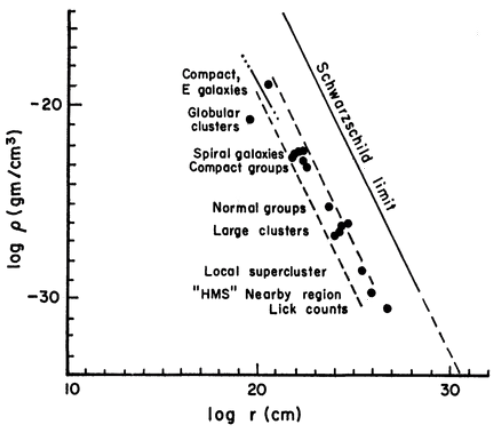}}
\caption{{Density-radius} 
 relation of various systems of galaxies \citep{de-Vaucouleurs:1970}.
 Reproduced with permission from AAAS, The Case for a Hierarchical
 Cosmology;  published by Science, 1970.}
\label{fig:DeVouc}
\end{figure}

In their studies, \citet{Carpenter:1938aa} and
\citet{de-Vaucouleurs:1970} observed that extragalactic entities
establish a linear correlation between their characteristic density
and radius when expressed in logarithmic terms, as~illustrated in
Figure~\ref{fig:DeVouc}. This correlation exhibits a slope of
approximately $-1.7$  and aligns with the Schwarzschild
limit. Furthermore, \mbox{\citet{de-Vaucouleurs:1970}} highlighted that
Abell's rich clusters are not only clustered on the characteristic
scale of superclusters but also extend to larger scales, indicating an
ongoing clustering phenomenon among galaxies. More recently,
\citet{Sankhyayan:2023aa} created a catalog of superclusters based
on galaxy clusters identified in the Sloan Digital Sky Survey (SDSS)
by \mbox{\citet{York:2000}.} The~authors derived the relationship between
density contrast and comoving size, discovering a slope of around $\sim$$-$2.

\begin{figure}[H]
\resizebox{0.49\textwidth}{!}{\includegraphics*{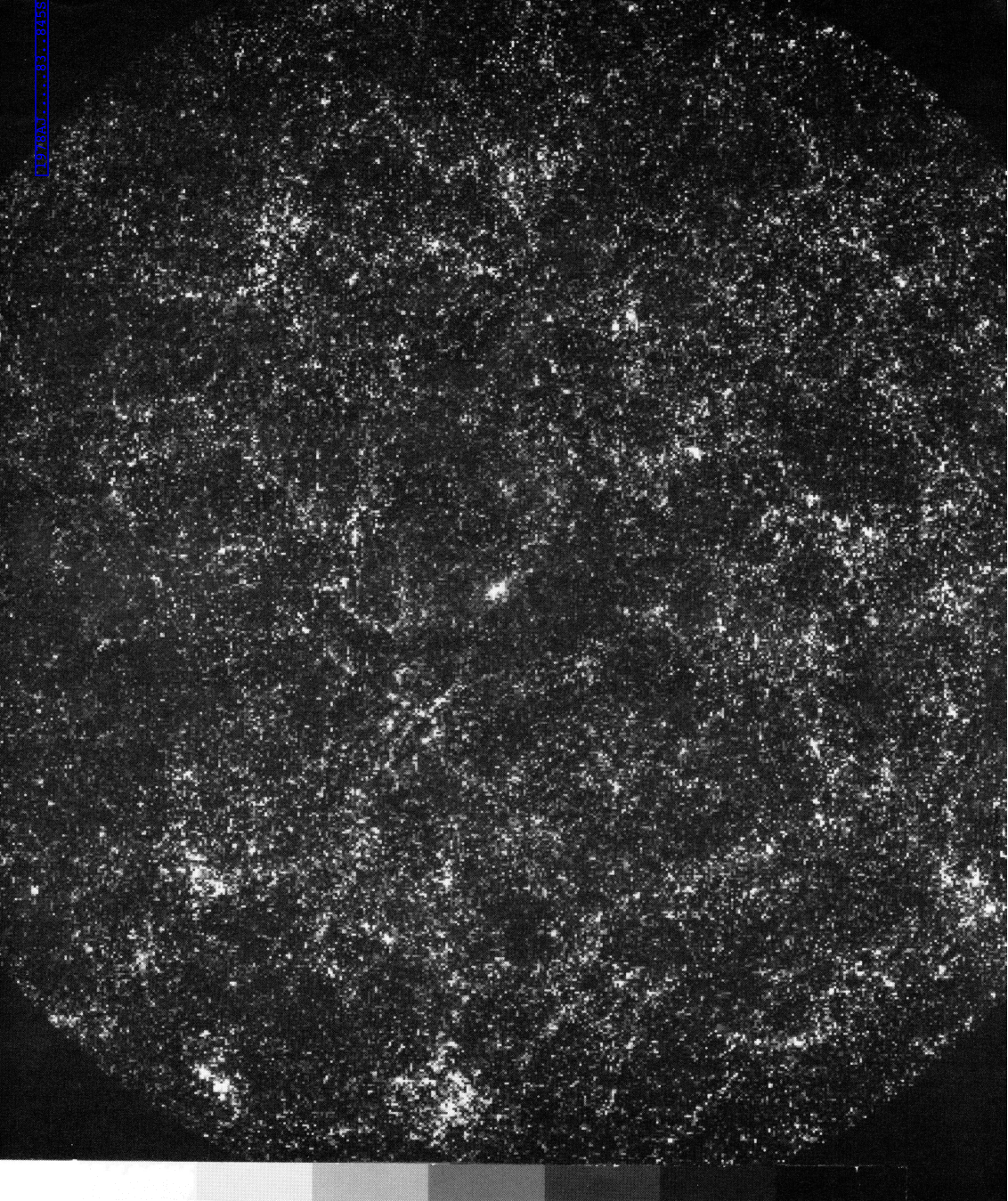}}
\resizebox{0.47\textwidth}{!}{\includegraphics*{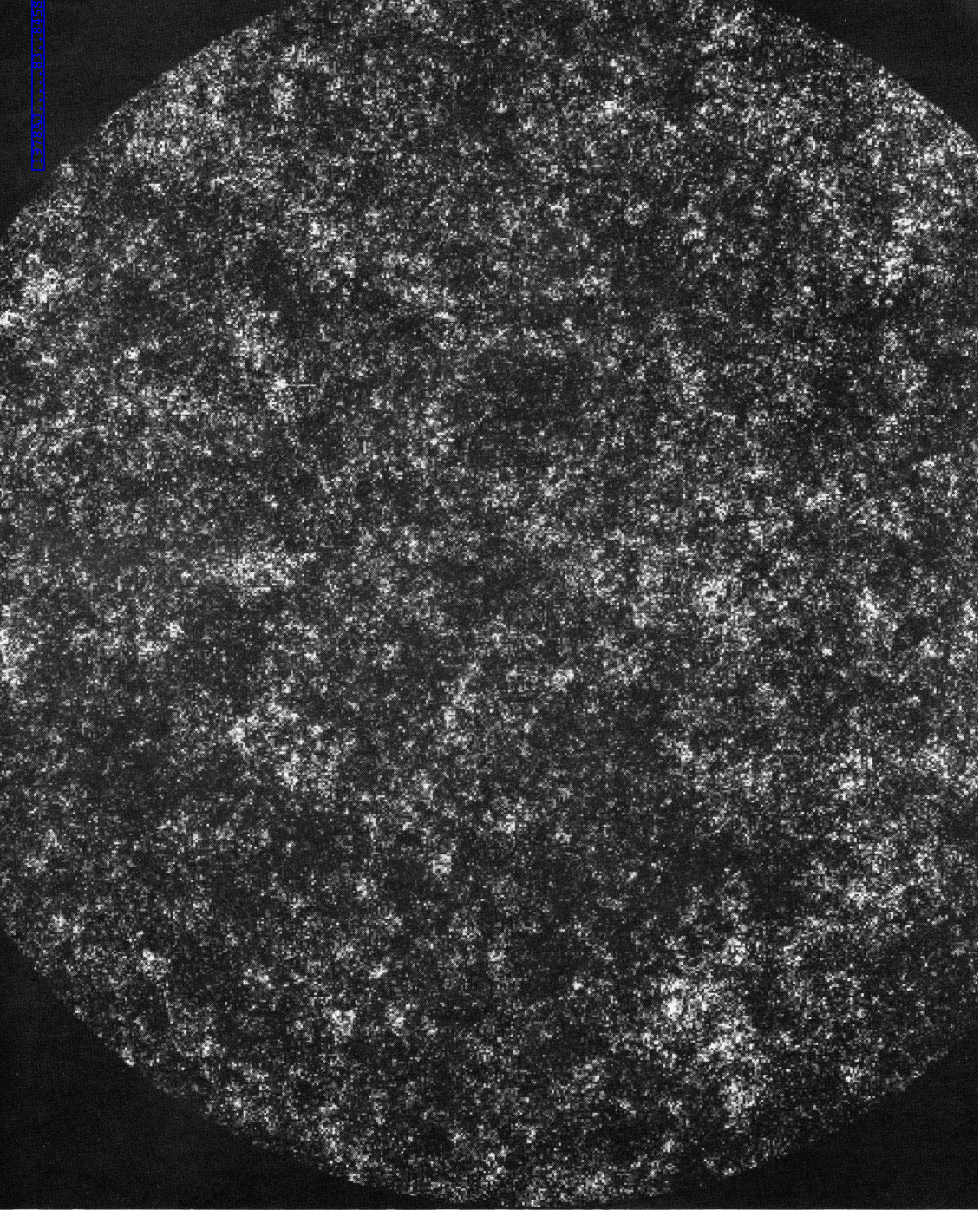}}
\caption{({\textbf{Left}):} 
 Map of Lick survey galaxies in the northern galactic hemisphere brighter
  than $m_B\le 18.9$ and north of galactic latitude $b\ge40^\circ$
  \citep{Soneira:1978fk}. (\textbf{Right}): Simulated map of galaxies
  imitating the 2D distribution of   Lick galaxies   \citep{Soneira:1978fk}. }
\label{fig:Lick}
\end{figure}

\textls[-25]{The first deep catalog of galaxies, covering the whole northern
hemisphere,  was made in the Lick Observatory with the 20-inch Carnegie
astrograph by \mbox{\citet{Shane:1967}.}  Actual counts of galaxies 
were made in $10'\times10'$ cells.  \citet{Seldner:1977a} used these
actual counts and~corrected the count for various errors and plate
sensitivity differences. The~final map of galaxies in the northern
galactic hemisphere $b\ge40^\circ$ is shown in the left panel of Figure~\ref{fig:Lick}.
Several well-known clusters of galaxies are seen on the map. For~example,
the Coma cluster appears near the center of the map. The~general
impression is that field galaxies are distributed approximately~randomly. }

\citet{Soneira:1978fk} developed a fractal 
model Universe to match the character of the galaxy distribution in
the Lick survey. The~model assigns `galaxy' positions in a
three-dimensional clustering hierarchy, fixes  absolute magnitudes,
and projects angular positions of objects brighter than $m=18.9$ onto
sky.  This procedure yields a galaxy map, shown in the right  panel of 
Figure~\ref{fig:Lick}.  Both the real Lick map and the computer
generated map were  used to calculate two-point angular correlation
functions.

\citet{Peebles:1973a} proposed utilizing the correlation method for
the analysis of distribution of galaxies, applying it to all
significant catalogs of extragalactic objects, including works by
\citet{Hauser:1973aa}, \citet{Peebles:1974ab}, \citet{Peebles:1974b},
  \citet{Peebles:1975}, and \citet{Peebles:1975a}.  These investigations demonstrated
that the angular distribution of galaxies could be characterized by a
power law. When the estimated angular correlation function is
converted to the spatial correlation function, it retains a power law
form:
\be
\xi(r)=(r/r_0)^{-\gamma},
\label{xilaw}
\ee
where $r_0 =4.5 \pm 0.5$~\Mpc\ is 
 the correlation length, and~$\gamma = 1.77$ is a characteristic power index \citep{Groth:1977aa}.
This power law is valid in the scale interval $0.05 \le r \le 9$~\Mpc,
where distances are expressed in units of the dimensionless Hubble
constant $h$ ($H=100~h$ km/s  per megaparsec).  \citet{Groth:1977aa}
showed that the angular correlation function is essentially zero at
angular distances $\theta \ge 10$ degrees.  The~conclusion from these
studies, based on the apparent (two-dimensional) distribution of
galaxies and clusters on the sky, confirmed the picture that galaxies
and clusters of galaxies are hierarchically~clustered.

\begin{figure}[H]
  \resizebox{0.55\textwidth}{!}{\includegraphics*{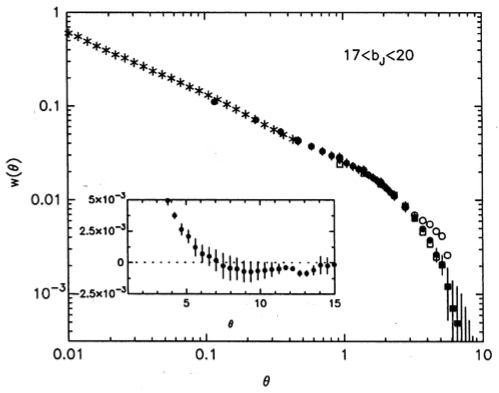}}
\caption{Average  angular correlation function of APM catalog of
  galaxies in the magnitude range $17 \le b_j \le 20$.  The~inset
  shows the mean angular CF on a linear scale. As argument, the
  angular separation in degrees is used   \citep{Maddox:1990aa}.}
\label{MaddoxFig}
\end{figure}

In 1970s and 1980s, British and Australian astronomers used Schmidt
telescope plates to photograph the whole sky.  The~Automatic Plate
Measuring (APM) machine in Cambridge was used to scan these
plates. Special software was developed to separate galaxy and star
images.  The~final catalog contains over two million galaxies
brighter than $b_j =20.5$.  \citet{Maddox:1990aa} used APM galaxies
to calculate the angular correlation function of galaxies, results are
shown in Figure~\ref{MaddoxFig}.  We see that the power law relation,
Equation~(\ref{xilaw}), is valid over three decades of angular distances,
$0.01 \le \theta \le 3$~degrees. The~almost constant slope of the
angular correlation function over a large range of angular scales was
interpreted by \citet{Peebles:2001aa} as  evidence that the spatial
correlation function is well represented by the law Equation~(\ref{xilaw})
over the range of separations 10 kpc$ \le r \le 10$~\Mpc.  As~we see
below in Section~\ref{2d3d}, this conclusion was
influenced by the insensitivity of the two-dimensional correlation
function to the spatial structure of the cosmic~web.

\subsection{Discovery of the Cosmic~Web}

\begin{figure}[H]
\resizebox{0.49\textwidth}{!}{\includegraphics*{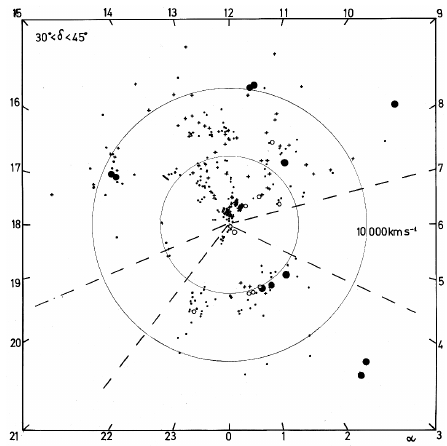}}
\resizebox{0.47\textwidth}{!}{\includegraphics*{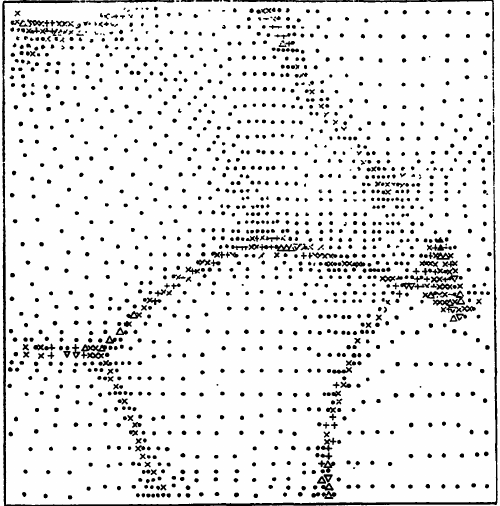}}
\caption{(\textbf{{Left }}:  Wedge diagram for the $30^\circ$ -- $45^\circ$ declination
  zone. Filled circles show rich clusters of galaxies, open
  circles---groups, dots---galaxies, crosses---Markarian 
  galaxies \citep{Joeveer:1978a}. (\textbf{Right}):~Distribution of particles in 
  simulations by Shandarin (1975,  private communication) and \citep{Doroshkevich:1980}.
}
\label{fig:wedges}
\end{figure}

In late 1970s, the number of galaxies with measured redshifts allowed
finding the distances of galaxies and studying the spatial
three-dimensional (3D)  distribution of
galaxies.  The first results of these analyses were reported in  the IAU
Symposium ``Large Scale Structure of the Universe'' in Tallinn,
September 1977  \citep{Longair:1978}.  \citet{Joeveer:1977py} presented the study of the
structure of the Perseus--Pisces Supercluster and its surroundings and
of the global network of superclusters and galaxy chains/filaments,
shown in Figure~\ref{fig:wedges}. 
Brent \citet{Tully:1978} presented a movie of the
Local Supercluster. To~obtain a spatial image of the supercluster, he
used the simple trick of making the image rotate, which created a
three-dimensional illusion. The~movie showed that the Local
Supercluster consists of a number of chains of galaxies that branch
off from the supercluster's central cluster in the Virgo constellation
as legs of a spider. William Tifft, in his talk, gave an overview of the recent study
of the Coma supercluster and its environ by \citet{Gregory:1978}.

The wedge diagram of galaxies in the $30^\circ$ -- $45^\circ$ declination
zone gives us a fascinating glimpse into the cosmic web. This
diagram reveals how galaxies within the Perseus--Pisces Supercluster
are arranged like a chain, with~clusters and groups of galaxies
appearing like pearls along a necklace. This structure is a prime
example of the cosmic web's basic elements: clusters, filaments,
sheets, and~voids.
Superclusters of galaxies are massive, but~they occupy only about 4\%
of the Universe's total space. The~remaining 96\% is composed of vast
voids. This large-scale geometry forms a continuous network that
includes clusters, filaments, sheets, and~the spaces between
them. Interestingly, central galaxies in rich clusters are typically
of the cD type and are often active radio~sources.

Sergei Shandarin's early numerical simulations were groundbreaking in
illustrating the evolution of particles through gravitational
clustering, based on the theory developed by \citet{Zeldovich:1970}. In~the
right panel of Figure~\ref{fig:wedges}, you can see a fascinating system
of high- and low-density regions. High-density areas are compact and
clumped together, forming a network of filaments that enclose
expansive under-dense regions. This visualization was pivotal, as it
gave the first glimpse into the Universe's structural patterns as
predicted by the Zeldovich~model.

The origin of this filamentary structure was analyzed by
\citet{Bond:1996}, who introduced the term ``cosmic web'' to
characterize this phenomenon. Analyses of galaxy distribution
indicated that the correlation length of clusters significantly
exceeds that of individual galaxies (see, for~instance,
\citet{Bahcall:1983uq, Klypin:1983fk}). This observation was
interpreted by \citet{Kaiser:1984} as a form of bias affecting
clusters in relation to galaxies. \citet{Szalay:1985aa} showed that,
if the correlation function is in the form Equation~(\ref{xilaw}), its
index $\gamma$ determines the fractal dimension of the sample:
$D= 3-\gamma = 1.23$.

\subsection{Discussion of the Fractal Character of the Cosmic~Web}

\begin{figure}[H]
\resizebox{0.50\textwidth}{!}{\includegraphics*{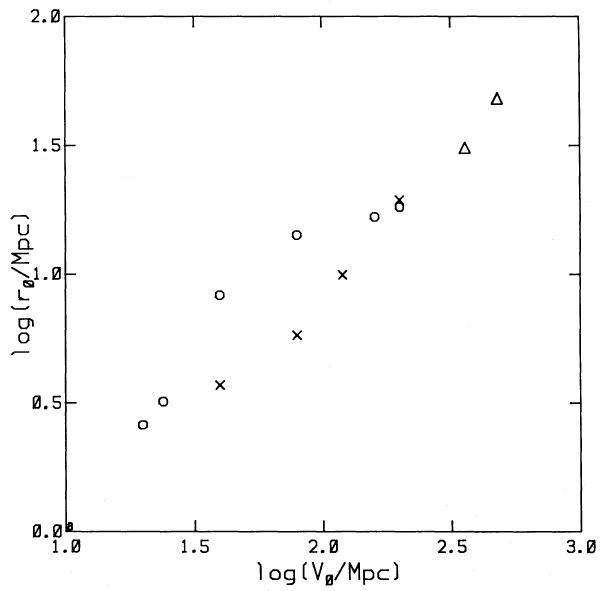}}
\caption{The correlation length, $r_0$ as function of the limiting
  redshift, $V_0$. Designations are as follows: open circles ---
  galaxy samples in the direction of the Coma supercluster; crosses ---
  galaxy samples in the direction of the Perseus-Pisces supercluster;
  triangles -- samples of rich clusters of galaxies.. Rich clusters, for~example, might show a
  larger correlation length compared to galaxies 
  \citep{Einasto:1986oh}. }
\label{KlypinFig}
\end{figure}

The next step was made by  \citet{Einasto:1986oh}, who  demonstrated that the
correlation length is influenced not only by the luminosity of
galaxies but also by the depth of the sample, as~illustrated in
Figure~\ref{KlypinFig}. This relationship between galaxy correlation
length and sample depth was interpreted by \citet{Pietronero:1987aa}
as evidence for a fractal structure in the distribution of
galaxies. Pietronero emphasized  that the fractal nature of galaxy
distribution extends to infinitely large distances, suggesting that
the entire Universe exhibits fractal characteristics. Furthermore,
\citet{Jones:1988nu} examined the galaxy distribution within the CfA
redshift survey and in a simulation of the $\Lambda$CDM model
conducted by \mbox{\citet{Gramann:1987ai,Gramann:1988},} which is recognized
as one of the first  $\Lambda$CDM simulations featuring
$64^3$~particles within a 40~\Mpc\ box. The~authors concluded that
both the 
galaxy distribution and the $\Lambda$CDM model can be effectively
described using a multifractal approach, indicating that the fractals
possess more than one scaling index. The~dimensionality of this
distribution varies between 1 and~3.

Fractal properties of the distribution of galaxies were discussed in
the IAU Symposium ``Large scale structures of the Universe'', held in
Balatonfured, Hungary, on~15--20 June 1987. Bernard Jones reported
basic results by \citet{Jones:1988nu}. He
started his talk showing the distribution of galaxies and declared
{\em ``this is a fractal''}.  The~distribution of both observed and model
samples can be described by multifractals with varying fractal
dimension. A~further discussion of the fractal character of the
large-scale distribution of galaxies was by \citet{Mandelbrot:1988aa},
who mentioned that he developed the multifractal concept long ago,
between 1970 and 1976, as described in his books
\citep{Mandelbrot:1977aa, Mandelbrot:1982uq}. Further, he concentrated
on the question: is the transition to homogeneity at the distance
$R_{cross}$ inside or outside of the limiting distance of data, 
$R_{max}$?  If~$R_{max} < R_{cross}$, we cannot make any decisions on
the transition scale to homogeneity.  After~the Symposium, Alex Szalay
invited a small group of interested people to Budapest to discuss in a
relaxed atmosphere the fractal character of galaxy distribution:
Benoit Mandelbrot, Yakov Zeldovich, Bernard Jones, and the author of the
present review.  In~the discussion, we agreed that the limit of the
validity of the power-law character of the correlation function with
constant index $\gamma = 1.77$ is at least  $2\,r_0\approx 10$~\Mpc.
Available data go beyond this distance and~show definitely a
multifractal character. However, inside the limit of observational and
model data, $R_{max} \approx 30$~\Mpc,  there is no evidence for the
transition to homogeneity with fractal index $D=3$.  Thus, further
studies are  needed to find the value of $R_{cross}$.  Zeldovich disliked
the fractal description of the Universe for two reasons: (i) it gives
no hint to the physics of the formation and evolution of the Universe,
and (ii) it contradicts other data that show that the mean density of
matter is not zero, as~predicted by a simple fractal~model.

Subsequent discussions of the fractal characteristics of the cosmic
web have been undertaken by various research groups, employing diverse
methodologies. The~majority of these discussions have centered around
the widely accepted concordant $\Lambda$CDM model. Key aspects of this
model were presented at several IAU Symposia: in Tallinn in
1977~\mbox{\citep{Longair:1978},} Crete in 1982 \citep{:1983aa},
Hungary in 1987 
\citep{:1988aa}, and~again in Tallinn in 2014
\citep{:2016aa}. The~theoretical underpinnings of this model are
rooted in the hierarchical 
clustering scenario proposed by \citet{Peebles:1970}, alongside the
pancake model for cosmic web formation introduced by
\citet{Zeldovich:1970}, and~its extension through catastrophe theory
as described by \mbox{\citet{Arnold:1982ab}.} Further advancements in
methodology involved the application of statistical measures to
investigate the fractal nature of galaxy distributions, as~explored by
\citet{Mandelbrot:1982uq} and \citet{Martinez:1990uq}. These methods
encompass various definitions of fractal dimensions, including the
Hausdorff dimension, capacity dimension, and~correlation dimension
(for definitions, refer to \citet{Martinez:2002}).

In late 1970s, a burst of interest in the fractal character of the
Universe emerged.  Different authors had various styles in fractal
studies: the Anglo-American style, Italian style, and~a more neutral
style, represented by Mandelbrot, Jones, Martinez, and
\citet{Balian:1988aa, Balian:1989aa}.  The latter authors confirmed the
bifractal character of galaxy distribution between scales from 0.1 to
10~\Mpc, with~different fractal dimensions for dense clustered
regions and~for underdense regions.  \citet{Song:1987aa} and 
  \citet{Ruffini:1988aa} constructed a cellular fractal model of the early
universe. The authors assumed that dark matter consisted of some 'inos',
which become non-relativistic at epoch $1+z_{nr}$, and~calculated the main
parameters of `` elementary cells'': characteristic Jeans masses are
$4\times 10^{17}~M_\odot$, radii 100~\Mpc, and~epochs
$1+z_{nr} \approx 10^4$.  The authors predicted that the cellular fractal
model has an upper cutoff, and~above this cutoff the mean density does
not decrease with distance.  This model was compared with observations
by \citet{Calzetti:1987aa, Calzetti:1988aa}. More recent
investigations into the fractal properties of the cosmic web have been
conducted by \citet{Gaite:1999aa} and \mbox{\citet{Gaite:2015aa, Gaite:2019aa}.}

The Anglo-American style of fractal studies was based essentially
on the angular distribution of galaxies. The first steps in this approach were
conducted by \mbox{\citet{Peebles:1970}} and
\mbox{\citet{Peebles:1973a},} who suggested 
the use of the correlation function to describe the distribution of
galaxies. The next essential step was the application of a fractal model
by \mbox{\citet{Soneira:1978fk}} to describe the angular distribution of
galaxies.  A further use of this approach was the study of the
distribution of APM galaxies by \citet{Maddox:1990aa}.  The~fractal
character of the cosmic web was analyzed by \citet{Peebles:1989aa} and
\citet {Peebles:1993aa}. It is characteristic that authors of the
Anglo-American style studies avoided in their publications the term
``cosmic~web''.

The Italian style of fractal studies is essentially the continuation
of earlier work by \citet{Charlier:1922aa},  \citet{Kiang:1967}, and
\citet{de-Vaucouleurs:1970, de-Vaucouleurs:1971aa} on the hierarchical
distribution of galaxies.  This style is represented by
\citet{Pietronero:1987aa,Pietronero:1997aa,Pietronero:2001aa},
\citet{Sylos-Labini:1996aa},  and \citet{Borgani:1995aa}.  The~focus
of Italian-style studies was the  fractal behavior of the Universe on
large~scales.

A dialogue between the Anglo-American and Italian views on fractal
properties of the Universe took place during the celebration of the
250th anniversary of Princeton University~\citep{Turok:1997aa}.  Marc
\citet{Davis:1997aa} presented the Anglo-American group's view.  His
main arguments were as follows: (i) the constant value of the correlation length
for 2D and 3D samples of various depths, $r_0\approx 4$~\Mpc; (ii)
the mean density of galaxies is the same for nearby and more distant
samples, and~the scatter of densities decreases with distance.
Luciano \citet{Pietronero:1987aa} described the Italian vision.
According to this view, the~correlation length of samples increases
with distance, and the mean density decreases with distance up to
$\sim$1,000~\Mpc.  Pietronero made a bet with Davis,  over~a case of
Italian or Californian wine, Neil Turok was the referee --- The
correlation length for volume-limited samples,  $M<-19.5$, is
$r_0\approx 5$~\Mpc\ (Davis) or $r_0\ge 50$~\Mpc\ (Pietronero). These
are fundamental questions, and I discuss these aspects of fractal studies
in later~sections.

\section{Statistics of Galaxy~Clustering  \label{spatial}}

Differences between fractal studies of various styles start from
variations in the methods to describe the fractal properties of the Universe.
The three-dimensional distribution of galaxies and clusters of
galaxies was described in the Tallinn 1977 symposium only
qualitatively.
In this section, I discuss some aspects of statistics related to the
estimation of quantitative statistical parameters of the cosmic~web.

\subsection{Measuring Spatial Distribution of~Galaxies \label{spatial2}}

In the late 1970s, astronomers and cosmologists began to realize that
the Universe's total density of matter was about 20\% of what we call
the ``critical density'' — the density needed for the Universe to be flat
and perfectly balanced.  Most intriguingly, they found that the
majority of this mass was in a mysterious form: dark matter. This was
highlighted by the pioneering work of \citet{Einasto:1974} and
\citet{Ostriker:1974}, who laid the groundwork for our understanding
of this cosmic puzzle (for a discussion see
\citet{de-Swart:2024aa}). Back then, scientists knew from Big Bang
nucleosynthesis (the process that created the first atomic nuclei)
that only about 5\% of this critical density was made up of baryonic
matter, which is the ``normal'' matter that makes up stars, planets,
and us. The rest had to be something else.  Given this gap,
the~scientific community began considering the possibility that dark 
matter was non-baryonic. The~first candidate was massive neutrinos,
which were then known as hot dark matter (HDM) because they moved at
relativistic~speeds.

The first quantitative comparison of Peebles' and Zeldovich's structure
formation models was conducted by \citet{Zeldovich:1982}. The authors
investigated the properties of the distribution of real galaxies in the
Virgo--Coma region using CfA data (sample O), the~distribution of particles in a
3D simulation by \citet{Klypin:1983zr}, calculated using the
assumption that the dark matter particle population is made of
neutrinos (sample A). The~second model H was constructed according to the
prescription described by \citet{Soneira:1978fk}.  The~two-dimensional
view of this model is shown in the right panel of Figure~\ref{fig:Lick}.
The authors used also the Poisson distribution of particles.  Three tests
were used: the spatial correlation function, percolation, and
multiplicity~tests.

\begin{figure}[H]

\resizebox{0.39\textwidth}{!}{\includegraphics*{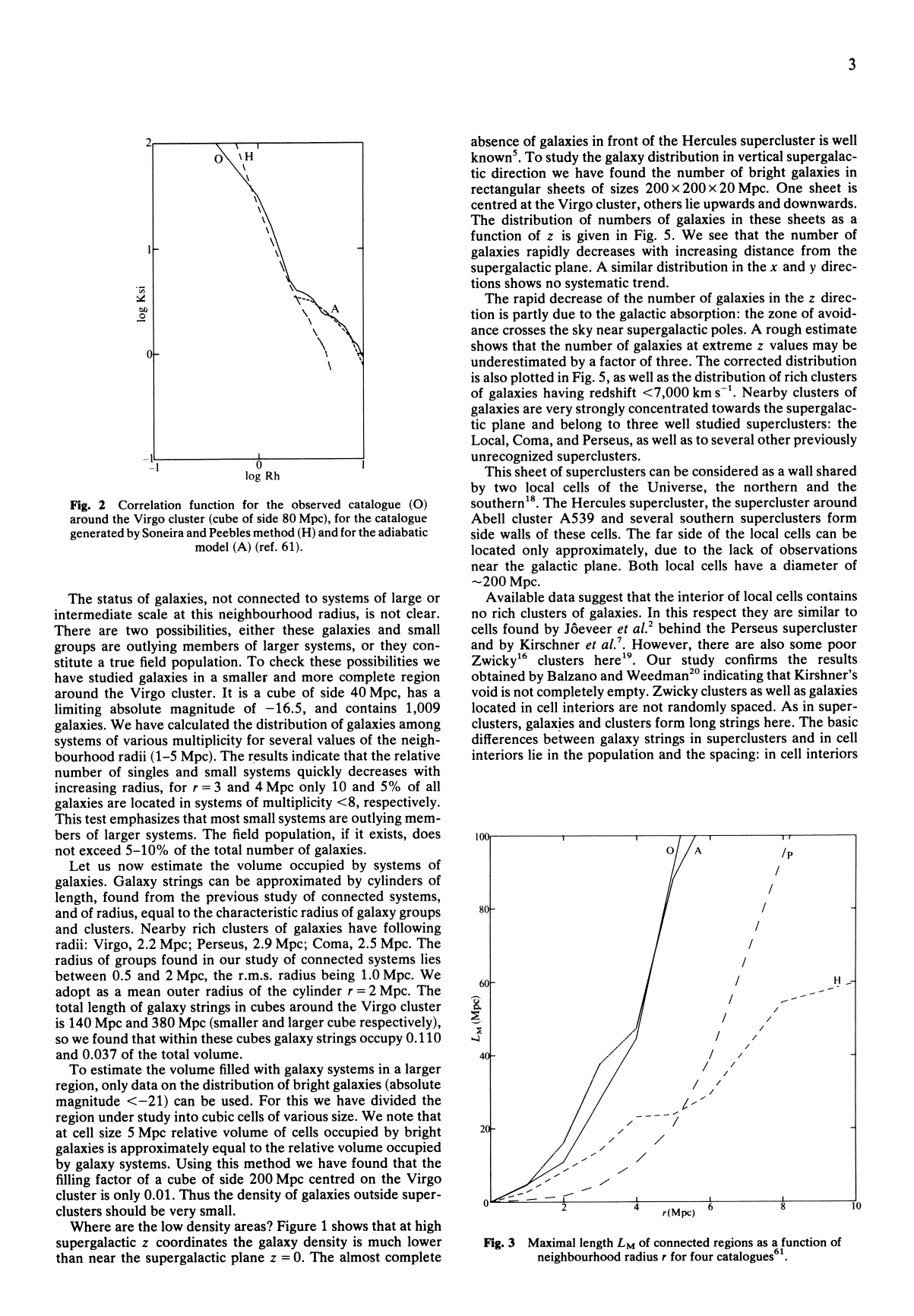}}
\resizebox{0.56\textwidth}{!}{\includegraphics*{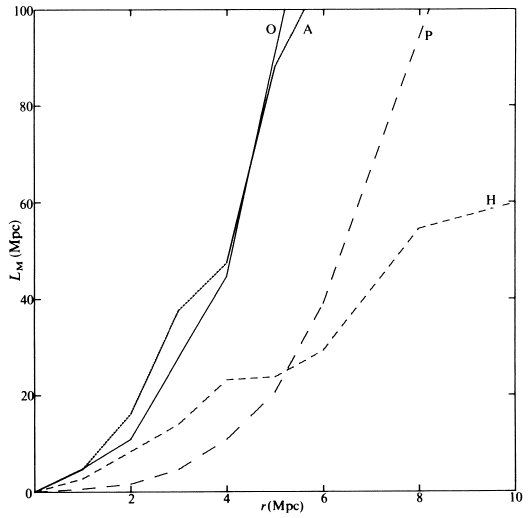}}
\caption{(\textbf{Left} {panel}): 
 the correlation function of the observed sample O
  around the Virgo cluster (cube of side 80 Mpc), of~the sample
  generated by the hierarchical clustering model H, and~of the
  adiabatic model A.  (\textbf{Right} panel): the maximal length $L_M$ of connected
  regions as a function of neighbourhood radius $r$ for four
  catalogs: O, A, H, and~P  (Poisson model). All distances are
  expressed for Hubble constant $h=0.5$ \citep{Zeldovich:1982}.
  Reproduced with permission from 
  Zeldovich, Y.B.;  Einasto, J.  Shandarin, S.F., Giant  voids in the universe,
  published by Nature 1982. 
  }
\label{fig:Lmax}
\end{figure}

The left panel of Figure~\ref{fig:Lmax} illustrates the spatial
correlation functions for three different samples. This figure
highlights a significant characteristic of the O and A samples: the
presence of a distinct knee in their correlation function, which is
notably absent in the hierarchical H and Poisson P models. At~short
distances, the~correlation function is highly sensitive to the
arrangement of galaxies or particles that are in close proximity to
one another. In~this range, most galaxies are found within clusters
and groups that typically exhibit an almost spherical
configuration. Conversely, at~greater distances, the~correlation
function reflects the existence of galaxy filaments that are primarily
one-dimensional in nature. Therefore, as~we transition from small to
large mutual distances among galaxies or particles, the~geometric
structure of the arrangement shifts. In~contrast, the~hierarchical and
Poisson models lack filaments, resulting in a correlation function
that appears~featureless.

The percolation method enables the assessment of the largest system's
length as a diagnostic tool. The~right panel of Figure~\ref{fig:Lmax}
illustrates the maximal lengths of galaxy and particle systems as a
function of the neighborhood radius $r$. The~neighborhood radius defines
a range of distances around a galaxy or particle, where other elements
are taken into account for analysis. Both galaxies and particles in
simulations exhibit clustering behavior; consequently, at~smaller
radii $r$, the~length of the longest system increases at a rate that
surpasses that of a Poisson sample. However, at~greater distances,
the~behavior of the samples diverges. In~both the observed sample O
and 
the model sample A, filaments connect clusters into a network. These
filaments facilitate the formation of longer systems, resulting in a
more rapid increase in the length of the longest system compared to
the Poisson scenario. As~depicted in Figure~\ref{fig:Lmax}, the~growth
patterns of samples O and A are nearly identical. In~contrast,
for~sample H, the~rate of increase of length $L$ with respect to
radius $r$ at 
larger distances is slower than that observed in the Poisson
sample. This can be attributed to the lower density of field particles
in sample H, as~a significant portion of the particles is concentrated
within clusters. Thus, this test proves to be sensitive to the
existence of filaments that link clusters into a cohesive~network.

The multiplicity test revealed distinct distributions of
multiplicities across all samples. The~observed sample exhibits a
relatively balanced representation of systems with varying richness,
indicating the presence of a detailed structure comprising galaxy
systems of diverse richness levels. Notably, the~majority of galaxies
are concentrated within a single extensive structure --- the Virgo
supercluster. In~contrast, the~A sample features a prominent large
system as well, but~its distribution of smaller systems closely
resembles that of a Poisson sample. This suggests a scarcity of
systems with intermediate richness, such as small-scale
filaments. Consequently, the~A sample, derived from the
neutrino-dominated Universe model, also appears to contradict the
observational~data.

The primary conclusion drawn from this analysis is that the
hierarchical clustering model proposed by \citet{Soneira:1978fk} fails
to perform adequately across all tests, while the adiabatic model
struggles specifically in the multiplicity test. Additionally, the~neutrino-based adiabatic model encounters a significant challenge: it
predicts that structure forms too late. Observational data indicate
that galaxies and rich clusters of galaxies formed earlier than this
model suggests, as~highlighted by
\citet{van-den-Bergh:1962}. Consequently, both conventional
neutrino-dominated cosmology and the hierarchical clustering model
exhibit shortcomings. To~address the challenges associated with
neutrinos as a candidate for dark matter, \citet{Peebles:1982aa}
proposed that dark matter consists of weakly interacting particles,
known as Cold Dark~Matter.

To evaluate the viability of the Cold Dark Matter (CDM) concept,
\citet{Melott:1983} conducted an analysis of the pioneering 3D CDM
simulation by \mbox{\citet{Centrella:1983mz}.} This examination revealed that the CDM
model aligns with all quantitative tests employed by \citet{Zeldovich:1982}. 
The formation of galaxies is initiated by the collapse of
small-scale perturbations, consistent with the clustering scenario
proposed by Peebles. In~contrast, large-scale structures develop in
accordance with Zeldovich's framework. The~concept of the cosmic web
was further refined by \citet{Bond:1996}, whose study elucidated the
mechanisms by which filaments are interconnected to create this
intricate network. However, both structure formation scenarios require
adjustments. Hierarchical clustering is not merely a random
occurrence; rather, it represents a continuous flow of particles and
galaxies directed toward attractors formed from the highest peaks of
the primordial fluctuation field. The~process of pancaking originates
from these peaks, resulting in various types of caustics, as~suggested
by \citet{Arnold:1982ab}.

\subsection{Measuring the Correlation~Function}

Early studies of the distribution of galaxies were based on
two-dimensional (2D) angular data as described in
Section~\ref{angular}.  To~measure the distribution in a quantitative way, the
correlation function was applied \citep{Peebles:1973a}.  As~discussed
by \citet{Peebles:1980aa},  the~angular correlation function of almost
all samples of galaxies is well represented by a power-law~function:
\be
w(\theta)=A\theta^{1-\gamma},
\label{xilaw1}
\ee
where $A$ is a constant, and~$\gamma$ is a parameter, whose value for
most samples studied was $\gamma\approx 1.7$.  This power law can be
inverted and has the solution:
\be
\xi(r)=Br^{-\gamma},
\label{xilaw2}
\ee
where $B$ is a constant, depending on $A$.  These equations show that
the angular correlation function is a power law with index lower by one unit than
the spatial correlation function~$\xi(r)$.

When, in the 1980s, galaxy samples with known radial velocities were
obtained, a~question emerged: how to use these three-dimensional (3D) data
to characterize the distribution in a quantitative way.
Distances of galaxies, calculated from observed radial velocities, are
influenced by the \citet{Kaiser:1987aa} effect --- an apparent
contraction of the galaxy density field in the radial direction.  To~avoid
this effect, \citet{Peebles:1973a} and \citet{Davis:1983ly} suggested
the use of the angular position of galaxies to find first the two-dimensional
correlation function.  In~this case, pair separations can be
calculated parallel to the line of sight, $\pi$, and~perpendicular to
the line of sight, $r_p$. The~angular correlation function,
$w_p(r_p)$, can be found by integrating over the measured
$\xi(r_p,\pi)$, using the equation
\begin{equation}
  w_p(r_p) = 2 \int_{r_{min}}^{r_{max}} \xi(r_p, \pi) d\pi,
  \label{wp}
\end{equation}
where $r_{min}$ and $r_{max}$ are the minimum and maximum distances of
the galaxies in the sample.  This equation has the form of the Abel
integral equation and~can be inverted to recover the spatial
correlation function \citep{Davis:1983ly}:
\begin{equation}
 \xi(r) = - {1 \over \pi} \int_r^{r_{max}} {w_p(r_p) \over \sqrt{r_p^2 -
     r^2}} d r_p.
 \label{xi}
\end{equation}

If the correlation function is described as a power law function,
angular and spatial functions have the forms of Equations~(\ref{xilaw1}) and
(\ref{xilaw2}), respectively. \citet{Davis:1983ly} made a 
correlation analysis of the CfA redshift survey with magnitude limit
14.5 and~applied the procedure described above to find correlation
function parameters. The authors  found that the spatial correlation function can
be well represented by a power law, Equation~(\ref{xilaw}), with parameters
$r_0=5.4 \pm 0.3$~\Mpc, and~$\gamma = 1.77$. 

In subsequent years, this procedure was applied in most correlation
analyses. \citet{Norberg:2001aa} investigated the luminosity
dependence of galaxy clustering in the 2dF Galaxy Redshift Survey. To~measure the correlation length, the authors used projected angular
correlation functions  as suggested by \citet{Davis:1983ly}.  The authors
found the real space correlation length $r_0=4.9\pm0.3$~\Mpc\ and
power law slope $\gamma=1.71\pm0.06$. \citet{Zehavi:2005aa,
  Zehavi:2011aa} studied the luminosity dependence of the SDSS galaxy
correlation function and~applied the standard procedure to measure
the projected correlation function.  Over~the scales $0.1 < r_p <
10$~\Mpc\, the power law approximation yields for correlation length of
$L^\star$ galaxies with $M_r=20.0$: $r_0=5.24\pm0.28$~\Mpc\, and
$\gamma=1.87\pm0.03$. 

The inversion
Equation~(\ref{xi}) assumes that spatial three-dimensional and projected
two-dimensional density fields are statistically similar. As~we see
below in Section~\ref{2d3d}, this assumption is not~valid.

\subsection{Measuring the Fractal~Dimension \label{dimens}}

The discovery of the dependence of the correlation length from the
type of objects by \citet{Bahcall:1983uq}, \citet{Klypin:1983fk}, and
\citet{Einasto:1986oh} and~the interpretation of this effect by
\citet{Pietronero:1987aa} in fractal terms initiated discussions on
the following topic: what are the best methods to characterize fractal
properties of the spatial distribution of galaxies?  This problem was
discussed also by \citet{Calzetti:1988aa}, \mbox{\citet{Coleman:1992aa}}, and
\citet{Borgani:1995aa},  who  pointed to the fact that, in the usual
correlation analysis, the observed galaxy distribution is normalized to
the Poissonian distribution in a way that cannot be used to test the
homogeneity of the~sample.

The natural estimator to determine the two-point  correlation
function is
\begin{equation}
  \xi_N(r) = {DD(r) \over RR(r)} - 1,
  \label{nat}
\end{equation}
where $r$ is the galaxy pair separation (distance), and~$DD(r)$ and
$RR(r)$ are normalized counts of galaxy--galaxy and random--random pairs
at a distance $r$ of the pair members.  Normalization equalizes the
sum of all $DD(r)$ to the sum of all $RR(r)$.  Galaxies are
clustered; thus, at small distances, the number density of galaxies is
enhanced, $DD(r) > RR(r)$, and~$\xi(r) >0$. At~large distances, the
density of galaxies is less than the mean galaxy density (a large
fraction of galaxies is located in clusters), thus $DD(r)< RR(r)$,
and~by construction at large distances $\xi(r) < 0$, as~found already
by \citet{Calzetti:1988aa}. The~relative volume 
of regions with $DD(r)< RR(r)$ is much larger than the relative volume
of regions with $DD(r) > RR(r)$; thus, the correlation function on
larger scales is only slightly negative, see Figure~\ref{MaddoxFig}.
The crossover at separation $r_c$, where $DD(r_c)=RR(r_c)$, is
approximately proportional to the depth of the~sample.

\citet{Pietronero:1987aa}, \citet{Calzetti:1988aa}, and
\citet{Coleman:1988aa} interpreted the increase in the galaxy
correlation length with the sample size with this normalization
effect and~suggested that, instead of $\xi(r)$, an alternative
clustering measure should be used: $\Gamma(r) = n\,(1+\xi(r))$, where
$n$ is the mean density of galaxies.  Another possibility is to use,
instead of the correlation function $\xi(r)$, the structure function
$g(r)=1+\xi(r)$, where $4\pi\,r^2g(r)n\dd{r}$ is the mean number of
galaxies lying in a shell of thickness $\dd{r}$ at distance $r$ from
any other point.  In~a Poisson process, $g(r)=1$. The~structure
function has a power law form on small scales, $r<5$~\Mpc,
and~approaches zero at large separations. In~the following analysis, I
use the structure function $g(r)$ to investigate fractal properties of the
distribution of~galaxies.

\section{Correlation Analysis of the Cosmic~Web  \label{cf}}

Historically, the~quantitative analysis of the cosmic web has been
dominated by correlation functions and their derivatives, the~structure
function and the fractal dimension function. It is well-known that the
correlation function contains information on amplitudes of the density
field but~not on their phases.  The~importance of the phase
information in the formation of the cosmic web has been understood
long ago. To~demonstrate the role of phase information, 
\citet{Coles:2000} extracted the simulated density field, Fourier
transformed the density field, and randomized phases of all Fourier
components.  The~modified field has on all wavenumbers $k$ the same
amplitudes as the original field, only the phases of waves are
different. In~the modified field, no structures are~visible.

Over the years, a variety of statistical methods have been developed to
analyze specific aspects of the spatial patterns in the large-scale
Universe.  Almost all these methods are borrowed from other branches of
science such as image processing, mathematical morphology,
computational geometry, and~medical imaging.  The~richness of various
methods to investigate the structure of the cosmic web is seen in
proceedings of the IAU Symposium ``The Zeldovich Universe: Genesis and
Growth of the Cosmic Web'' \citep{:2016aa}.

In the early years of the 21th century, new redshift surveys were
published---the 2dF and the Sloan Digital Sky Survey (SDSS).  The~2dF
survey by \citet{Colless:2003aa} allowed finding angular correlation
functions of 2dF galaxies \citep{Maddox:1990aa}, discussed in
Section~\ref{angular}. Next, the SDSS was available \citep{York:2000,
  Aihara:2011aa, Alam:2015aa}, which allowed studying the distribution
of galaxies in much larger volumes of space.  It was used  to calculate
correlation functions and power spectra of SDSS galaxies by 
\citet{Tegmark:2004aa} and \citet{Zehavi:2011aa}.  New large
numerical simulations of the cosmic web were developed, which included
hydrodynamical processes of formation and evolution of galaxies ---
the Millennium simulation by \citet{Springel:2005} and the Illustris
The Next Generation (IllustrisTNG) simulation by \mbox{\citet{Springel:2018aa}.}
These observational and modeling possibilities allowed studying the
character of the distribution of dark matter and galaxies in much
more~detail.

As described above, various groups obtained very different pictures of the
fractal characteristics of the cosmic web. Thus, it is evident that a
new independent study is needed, using more recent observational data
and simulations. This was conducted by \mbox{\citet{Einasto:2020aa, Einasto:2021ti}.}
In this section, I describe the conventional correlation analysis of
the cosmic web, using as tests SDSS galaxies and~particles from
several modern $\Lambda$CDM model 
simulations. First, I discuss the formation of galaxies in the cosmic
web and the method, how to select particles in simulations to form
samples of particles, comparable to samples of~galaxies.

\subsection{Formation of Galaxies in the Cosmic~Web}

By comparing spatial distributions of dark matter particles and
galaxies, \mbox{\citet{Joeveer:1977py}} and \mbox{\citet{Zeldovich:1982}} found that
there are almost no galaxies in voids, but~voids are populated by a
rarefied field of DM particles, as shown by simulation by Shandarin
and \mbox{\citet{Klypin:1983zr}}, 
see Figure~\ref{fig:wedges}. The authors of both papers 
emphasized from this difference that the galaxy formation is a
threshold phenomenon.  The~analysis by \citet{White:1978} confirmed
this: the galaxy formation is a two-stage process: first, dark matter
condenses to form heavy halos, where various hydrodynamical processes
form visible galaxies.  The first numerical simulations of galaxy
formation with a hydrodynamical method by \citet{Cen:1992fr} confirmed
this model, verified by \citet{Springel:2018aa} by a much more
detailed hydrodynamical simulation.  In~high-density regions the
baryonic matter forms galaxies, and in~low-density regions it remains in
the pre-galactic diffuse form together with low-density field of dark~matter.

Based on these arguments, it is natural to use particles in
high-density regions to get a sample of DM particles that imitates
samples of galaxies.  We  apply a sharp particle density
limit, $\rho_0$, to~select biased samples of particles.  This method
is similar to the Ising model, discussed by \citet{Repp:2019wl}.
Actually galaxy formation is a stochastic process; thus, the matter
density limit, which divides unclustered and clustered matter, is
fuzzy. However, a~fuzzy density limit has little influence on the
properties of correlation functions of biased and non-biased samples.
Thus, we can accept a fixed threshold limit and~select for biased
model samples particles with density labels, $\rho \ge \rho_0$.

\subsection{Correlation Functions of Galaxies and~Matter}
  
In early studies of the spatial distribution of galaxies, only samples
with a rather low distance limit were available, which raised the question: how
representative are these samples in terms of  describing the whole cosmic web?  As
discussed in Section~\ref{history}, various authors interpreted these
early data in a very different way.  To~avoid these difficulties, I use
in the following analysis only galaxy and model samples found in a
large sample volume, as~conducted  by \citet{Einasto:2020aa, Einasto:2021ti}.

\begin{figure}[H]
 \hspace{-6pt}
\resizebox{0.98\textwidth}{!}{\includegraphics*{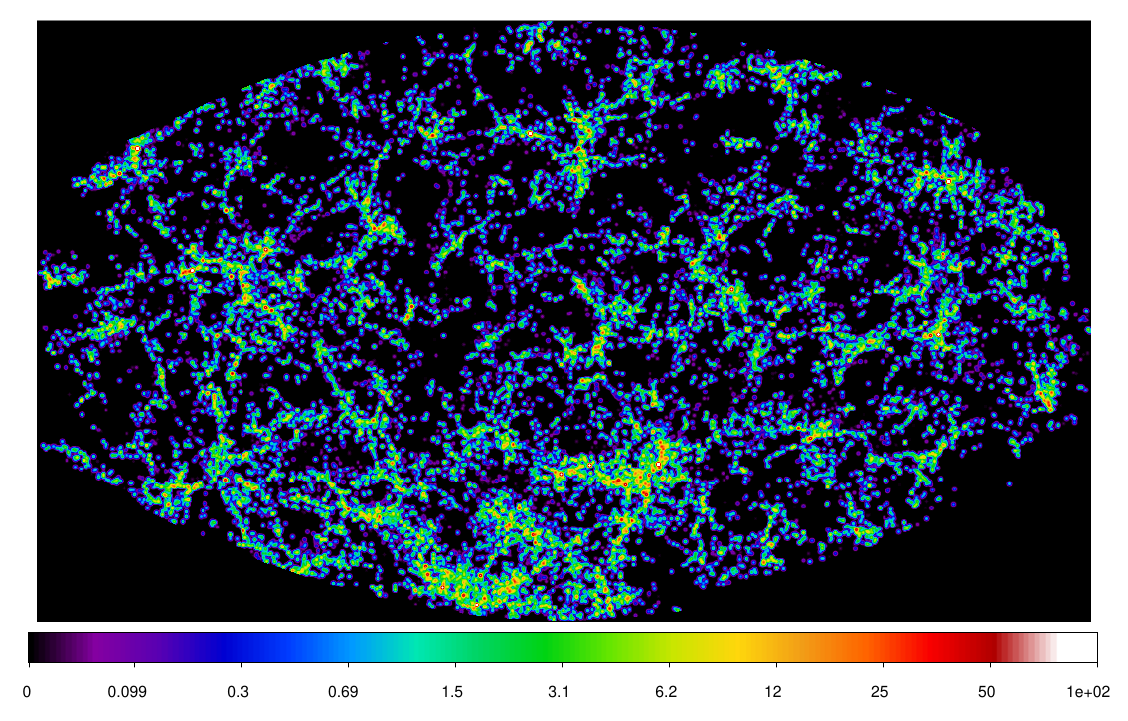}}
\caption{Slice  of the density field from the Sloan Digital Sky Survey
at a distance of $240$~\Mpc\ and thickness of $10$~\Mpc.
At lower part of the figure, the Sloan Great Wall is seen 
\citep{Einasto:2024aa}. Reproduced with permission from
    Einasto, J., Dark Matter and Cosmic Web Story;  published by World Scientific, 2024.  
}
\label{SDSSslice} 
\end{figure}

\citet{Einasto:2020aa} used the luminosity-limited galaxy
samples by \citet{Tempel:2014aa}, selected from data release 10 
of the SDSS galaxy redshift survey \citep{Ahn:2014aa}.  The~catalog
has a Petrosian $r${-}band magnitude limit $m_r \le 17.77$
and~contains 489,510 galaxies. The~SDSS samples have 
$M_r - 5\log h $ magnitude limits $-18.0$, $-19.0$, $-20.0$, 
$-21.0$, and~$-22.0$ and~ are referred to as SDSS.18t, SDSS.19t, SDSS.20t,
SDSS.21t, and~SDSS.22t.  The~effective size of the sample is 500~\Mpc.
One view of the SDSS density
field is presented in Figure~\ref{SDSSslice}. We see here a complicated
network of clusters, filaments, and voids. The~rich complex of
superclusters in the lower part of the Figure is the Sloan Great Wall,
which actually consists of three superclusters \citep{Liivamagi:2012}.

\vspace{-6pt}

 To~have both high spatial resolution
and the presence of density perturbations in a large scale interval,
\citet{Einasto:2020aa} used a series of simulations of the
$\Lambda$CDM models with box sizes $L_0=256,~512,~1024$~\Mpc\ with
$N_{\mathrm{grid}} = 512$ and~number of particles
$N_{\mathrm{part}} = 512^3$.  The cosmological parameters for all
simulations are ($\Omega_m,\Omega_{\Lambda},\Omega_b,h,\sigma_8,n_s$)
=(0.28,~0.72,~0.044,~0.693,~0.84,~1.00). In~the present analysis, I use
the model of size 512~\Mpc.  Additionally, I use the simulated galaxy
sample of the Millennium simulation by \citep{Springel:2005} and
\citet{Croton:2006}, which has the box of size 500~\Mpc, and~the EAGLE
simulation by \citet{McAlpine:2016aa}.  EAGLE simulations were run in
boxes of sizes 25, 50, and 100~\Mpc.

In the \citet{Einasto:2020aa} simulation, the~authors calculated local
density values, $\rho$, at~particle locations using the locations of
the 27 nearest particles, and~expressed the densities in units of the 
average density.  The~authors formed samples corresponding to the
simulated galaxies, containing particles that exceeded a certain
density limit, $\rho \ge \rho_0$.  These samples are denoted as
LCDM.$i$, where $i$ denotes the particle density limit $\rho_0$.
The~full DM model covers all particles, corresponds to the particle 
density limit $\rho_0 = 0$, and is therefore denoted~LCDM.00.

\begin{figure}[H]
\resizebox{0.48\textwidth}{!}{\includegraphics*{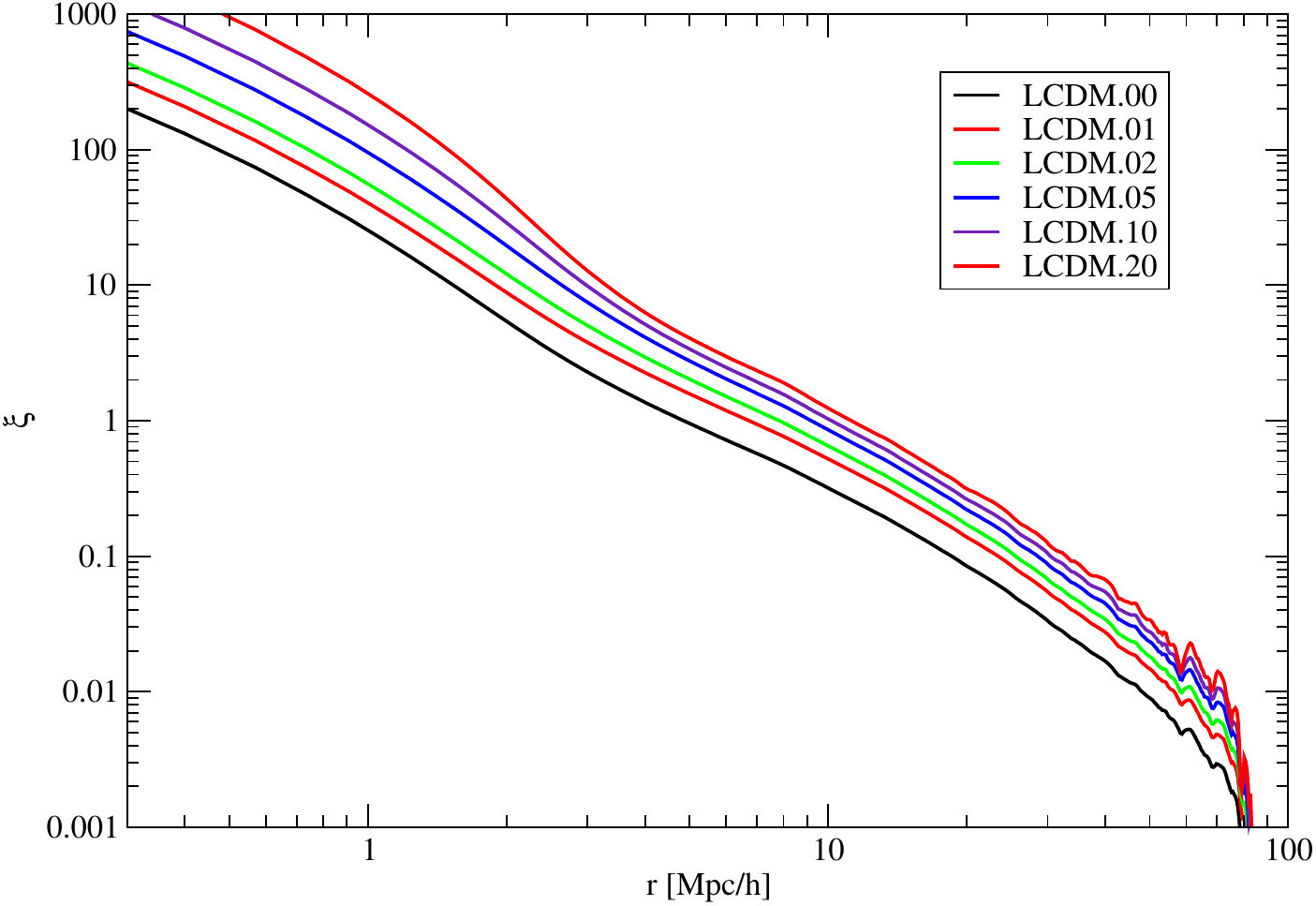}}
\resizebox{0.48\textwidth}{!}{\includegraphics*{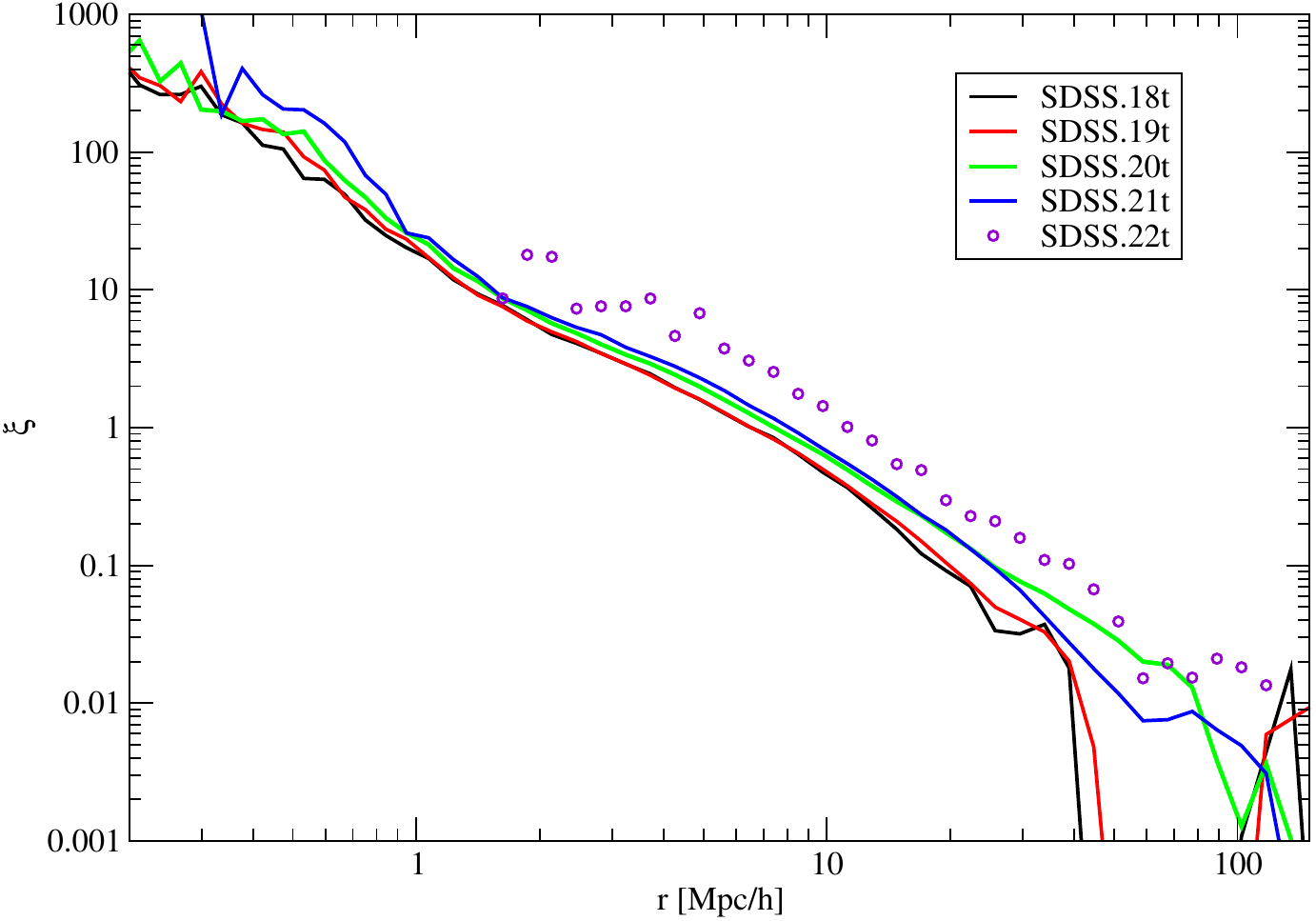}}
\caption{Correlation  functions of galaxies, $\xi(r)$. (\textbf{Left} panel) shows
  $\Lambda$CDM model of box size 512~\Mpc\ for different particle selection
  limits. (\textbf{Right} panel) is for SDSS galaxies using five luminosity
  thresholds \citep{Einasto:2020aa}.}
\label{fig:corrFig1} 
\end{figure}

\textls[-15]{Correlation functions of $\Lambda$CDM and SDSS samples are 
shown in the left and right panels of  Figure~\ref{fig:corrFig1}, respectively.
Figure~\ref{fig:corrFig1} shows that samples with different galaxy
luminosity and particle density limit form approximately parallel
sequences, where the amplitude of the correlation functions increases with
the increase in the luminosity/density  limit.  For~galaxy samples, the~amplitudes of the
correlation functions are almost constant for low-luminosity samples
and rise for samples brighter than approximately $M_r=-20$. The~behavior of $\Lambda$CDM model particle density selected correlation
functions is different---with increasing particle density threshold
$\rho_0$, the~amplitudes rise continuously.  The~luminosity dependence
of the correlation functions is the principal factor of the biasing
phenomenon, as~shown  by \citet{Kaiser:1984}. A~further discussion
of the correlation length is given in the following~subsection. }

\subsection{Luminosity Dependence of the Correlation~Length \label{luminosity}}

The dependence of the correlation length on the size and luminosity of
samples was the main object of early analyses by
\citet{Pietronero:1987aa}, \citet{Pietronero:1997aa}, and
\citet{Davis:1997aa}.  \citet{Pietronero:1997aa} defended the view
that the correlation length increases with the size of samples until
very large scales, $r\approx 1000$~\Mpc.  \citet{Davis:1997aa} argued
that it remains constant, $r_0\approx 5$~\Mpc, for all sample sizes.
I use new analyses of the correlation length to have a fresh view of
the~problem.

\begin{figure}[H] 
\resizebox{0.65\textwidth}{!}{\includegraphics*{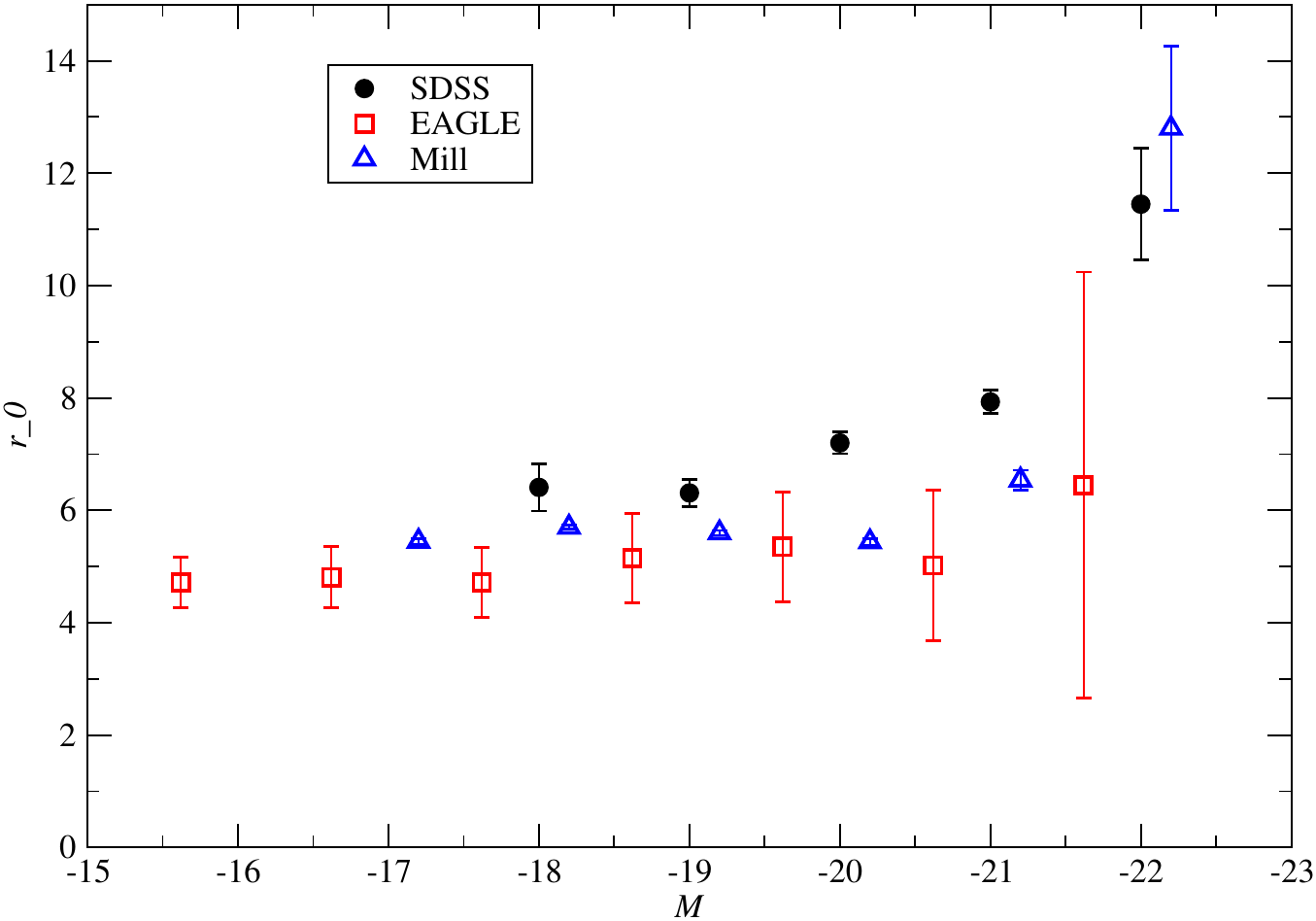}}
\caption{Correlation  length $r_0$ in \Mpc\ of SDSS galaxies as a
  function of their absolute magnitudes. For~comparison, we
  also show the correlation lengths of the EAGLE and Millennium
  simulations for various magnitude bins  \citep{Einasto:2020aa}. } 
\label{fig:corrFig6} 
\end{figure}

The data presented above allow calculating the correlation lengths of
observed SDSS samples and simulated Millennium and EAGLE samples.
In~Figure~\ref{fig:corrFig6},  I show the correlation lengths, $r_0$, of~the
SDSS, EAGLE, and Millennium samples as functions of magnitude $M_r$.
We see that, for low and intermediate luminosities, all samples have
correlation lengths $r_0\approx 5$~\Mpc, which rises to higher values
for more luminous galaxies. The~correlation length $r_0$ has a rather
similar luminosity dependence for all samples.  The~luminosity
dependence of the correlation functions is the principal factor of the
biasing phenomenon, as~discussed by \citet{Kaiser:1984}.

SDSS and Millennium samples have sizes 500~\Mpc, which is sufficiently
large to consider them as  representative for the whole cosmic web. Our analysis
shows that correlation lengths of low and medium luminosity galaxies
have the value $r_0\approx 5$~\Mpc, as~predicted by
\citet{Davis:1997aa}, and~contrary to the prediction by
\citet{Pietronero:1997aa}.  But~notice that the correlation length of
the EAGLE sample at low luminosities, $r_0\approx 4.5$~\Mpc, is
smaller than for the SDSS and Millennium samples.  This can be the sample
volume effect as discussed by \citet{Pietronero:1997aa}, since the
size of EAGLE samples is smaller than sizes of SDSS and Millennium~samples.

\textls[-25]{A significant aspect of luminosity dependence is the observation that
the correlation length remains nearly constant at low luminosities,
 $M \ge -20.0$. This phenomenon is evident in the
SDSS.19 and SDSS.18 observational samples; however, it cannot be
extrapolated to lower luminosities due to the lack of very faint
galaxies in the luminosity-limited samples from SDSS. In~the galaxy
samples from the EAGLE and Millennium models, a~gradual decline in
$r_0$ with decreasing luminosity can be tracked down to very faint
galaxies, with~$M \approx -15.6$ in the EAGLE sample and $M \approx
-17.2$ in the Millennium sample. Similar findings have been reported
by \citet{Norberg:2001aa} and \citet{Zehavi:2011aa}, indicating that
the correlation lengths of low luminosity galaxies approach a specific
limit as luminosity decreases. This trend suggests that very faint
galaxies tend to follow the spatial distribution of their brighter
counterparts, implying that faint galaxies often serve as satellites
of more luminous~galaxies.}

\section{Fractal  Analysis of the Cosmic~Web \label{web}}

To describe fractal properties of the cosmic web, most authors applied
the correlation function and its derivatives, the~structure function
and the fractal dimension function.  In~this section, I describe how
these functions can be used to analyze fractal properties of the
cosmic web.  In the fractal analysis, I use the same set of SDSS
and model samples as discussed in the previous section.  Model and
SDSS samples have almost identical volumes; thus, the volume dependence
of fractal properties is absent, and~we see the luminosity (particle
density limit) dependence of~samples.

The natural estimator to determine the two-point spatial correlation
function is given by the function Equation~(\ref{nat}).  Based on arguments
discussed in Section~\ref{dimens}, I use  in the following
analysis the structure function,
\be
g(r)=1 +\xi(r),
\label{struct}
\ee
and its log--log gradient, the~gradient function,
\begin{equation}
  \gamma(r)= {d \log g(r) \over d \log r},
  \label{gamma}
\end{equation}
which I call  the $\gamma(r)$ function.

\citet{Martinez:2002} defined the correlation dimension
\be
D_2 = 3 + d \log \hat{g}(r) / d \log r,
\ee
where $\hat{g}(r)$ is the
average of the structure function,
\be
\hat{g}(r) = 1/V \int_0^r{g(r`) d  V}.
\ee
The  parameter $D_2$ is related to the
effective fractal dimension function $D(r)$ of samples at mean
separation of galaxies at $r$ \citep{Pietronero:1987aa,
  Martinez:1990uq},

\textls[-25]{For our study, I prefer to use the local value of the structure
function to define its~gradient:}
\begin{equation}
 D(r) =3+ \gamma(r).
  \label{dim}
\end{equation}
Notice that the fractal dimension is defined for a range of scales
$r$; thus, the definition (\ref{dim}) is only an approximation of the true
fractal dimension. Also notice that  the $\gamma(r)$ function has the
opposite sign compared to the parameter $\gamma$ in the correlation
function, Equation~(\ref{xilaw}). 

\begin{figure}[H]
\resizebox{0.48\textwidth}{!}{\includegraphics*{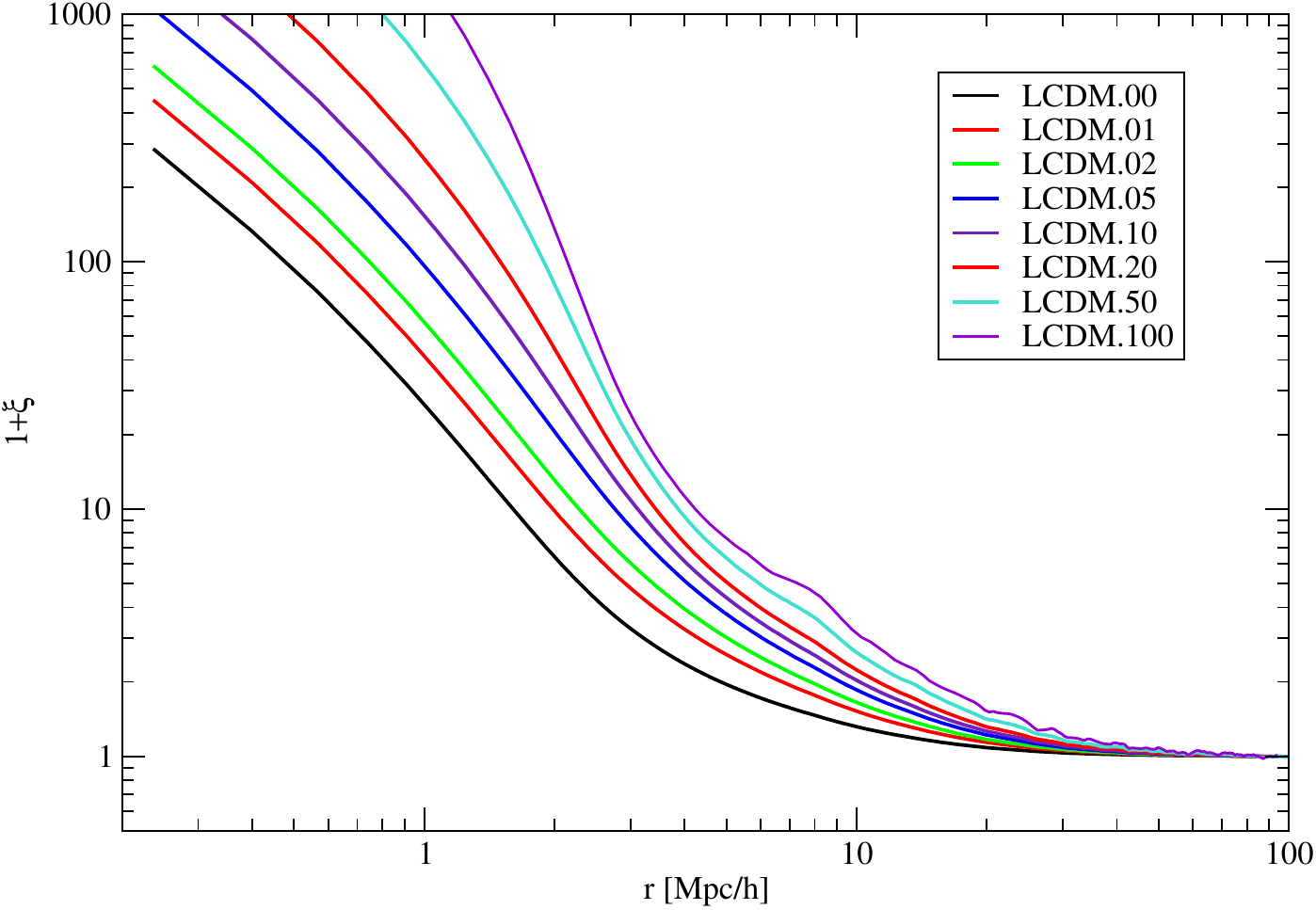}}
\resizebox{0.48\textwidth}{!}{\includegraphics*{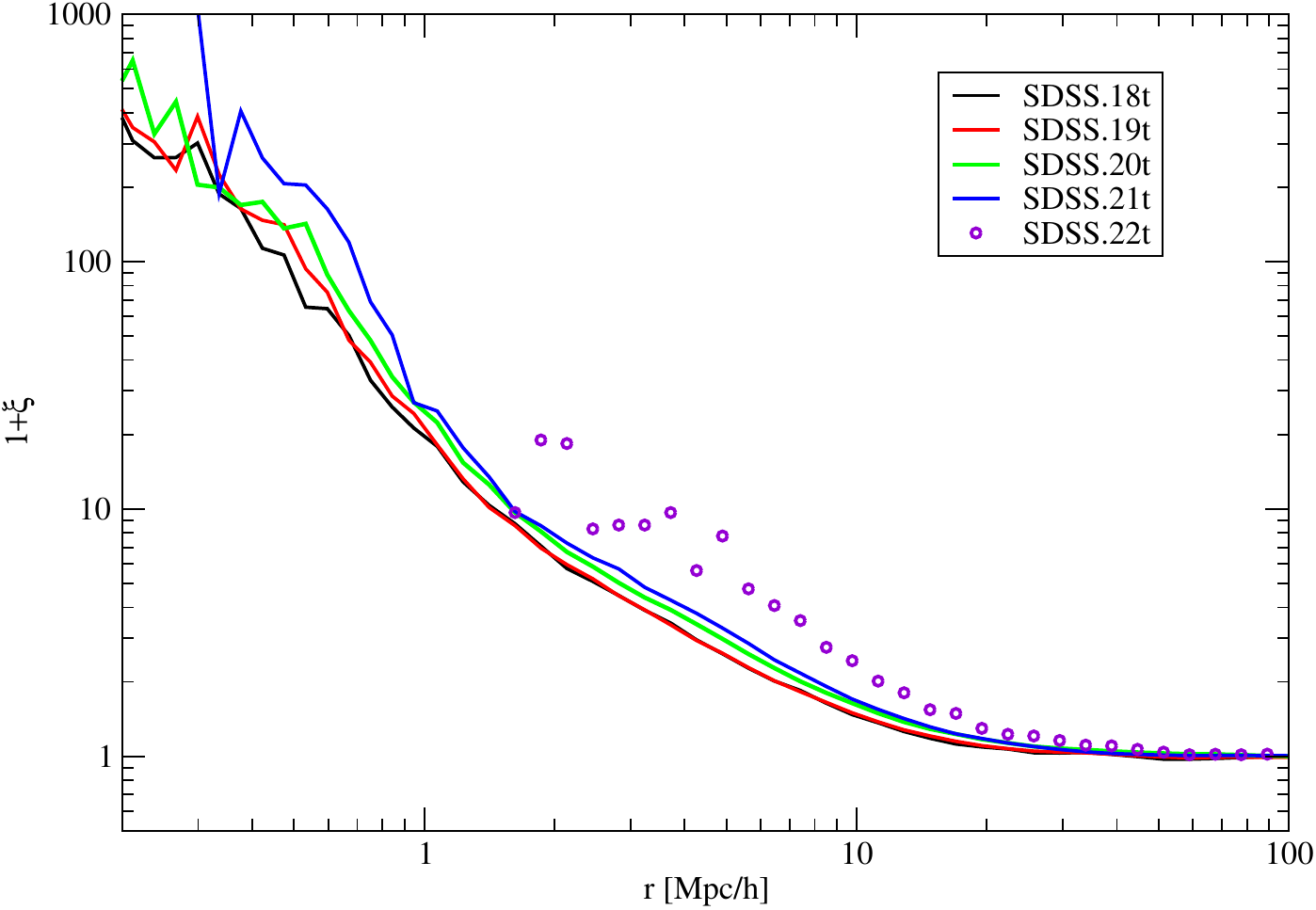}}
\caption{The  structural functions are defined as
  $g(r)=1+\xi(r)$. The~ positioning of the panels remains consistent
  with what is   illustrated in Figure~\ref{fig:corrFig1}
  \citep{Einasto:2020aa}. }
\label{fig:corrFig2} 
\end{figure}

\begin{figure}[H]
 
\resizebox{0.48\textwidth}{!}{\includegraphics*{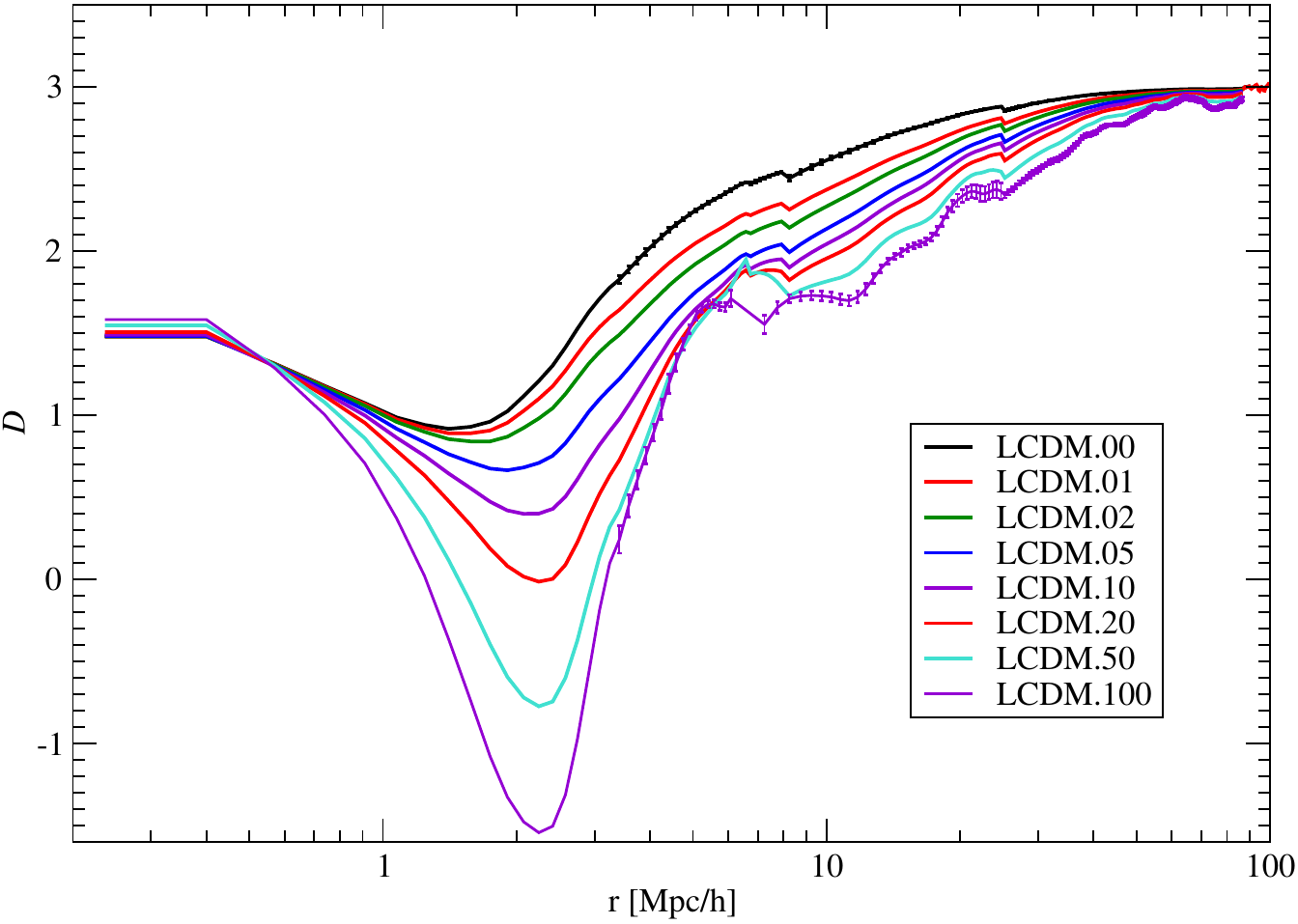}}
\resizebox{0.48\textwidth}{!}{\includegraphics*{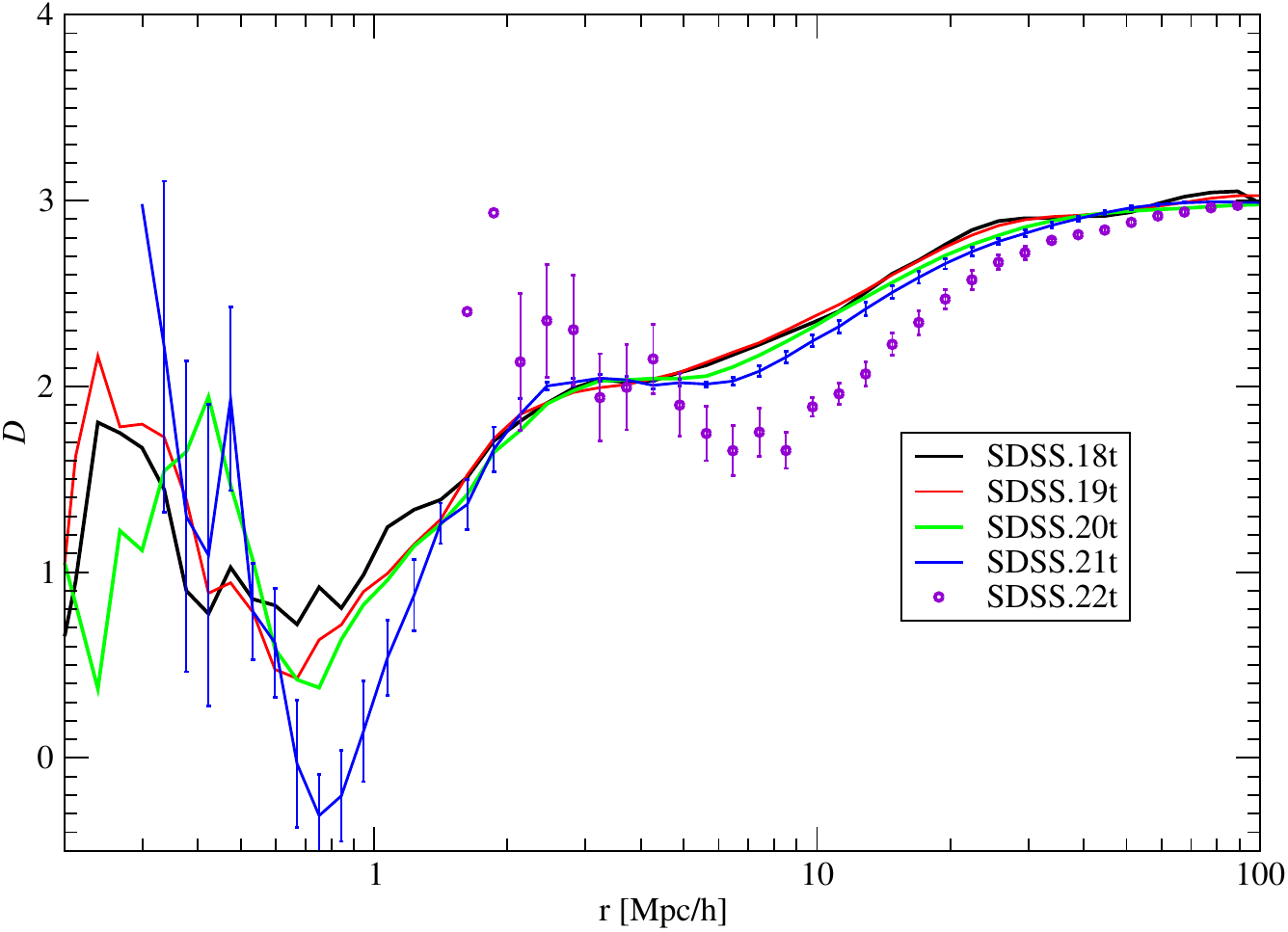}}
\caption{The  fractal dimension functions are expressed as
  $D(r)=3+\gamma(r) $. The~positioning of the panels corresponds to
  that depicted in Figure~\ref{fig:corrFig1}. Error values are provided
  for a selection of representative samples
  \citep{Einasto:2020aa}.} 
\label{fig:corrFig3} 
\end{figure}

In the previous section, we examined the correlation functions of our
model alongside the observed samples. Figure~\ref{fig:corrFig2}
illustrates the structure functions, represented as \linebreak $g(r)=1 + \xi(r)$, while
Figure~\ref{fig:corrFig3} depicts the fractal dimension functions,
denoted as $D(r)=3+\gamma(r)$. Notably, the~last  figure clearly indicates
that the fractal dimension function features two distinct regions,
with a transition occurring at a separation of approximately $r \approx3$~\Mpc.
This phenomenon has been previously identified by
\mbox{\citet{Zeldovich:1982}} and \mbox{\citet{Zehavi:2004aa}.} In~the case of
smaller mutual separations $r$, the~correlation function effectively
describes the distribution of matter within dark matter halos,
whereas, at~larger separations, it pertains to the distribution of the
halos~themselves.

\unskip

The fractal dimension function is a crucial concept in understanding
the geometric properties of complex structures in cosmology,
particularly in analyzing the distribution of particles in the
Universe. In~the context of $\Lambda$CDM (Lambda Cold Dark Matter)
samples, this function provides insights into how matter is
distributed on various scales. The~parameter $\rho_0$ represents a
particle density limit, which is essential when selecting particles
for analysis. By~adjusting this limit, researchers can examine how
different densities affect the fractal characteristics of the
sample.
The left panel of Figure~\ref{fig:corrFig3} illustrates that the fractal
dimension function for the $\Lambda$CDM samples is influenced by the
particle density threshold, $\rho_0$, employed in the selection of
particles for the sample. All $\Lambda$CDM samples exhibit a uniform
gradient function value of $\gamma(0.5) = -1.5$ at a distance of
$r=0.5$~\Mpc, corresponding to a local fractal dimension of $D(0.5)=1.5$. At~approximately 2~\Mpc, the~gradients reach a minimum that varies
according to the particle density limit $\rho_0$ of the samples.
The observed minimum in the gradients
at around 2~\Mpc\ suggests a transition in the distribution of
particles. This may imply that there are significant changes in the
spatial arrangement of matter at this scale, reflecting the
complexities of cosmic structures. After~this point, the~increase in
the fractal dimension function signifies a return to a more regular
distribution, culminating in the expected maximum value of $D(100) =
3.0$ at the largest distances. This value signifies a uniform and
isotropic distribution of matter, consistent with the assumptions of
the $\Lambda$CDM model at cosmological~scales.

The equivalent values of the fractal dimension functions at $r$ = 0.5~\Mpc,
succeeded by a minimum around  $r \approx 2$~\Mpc, can be attributed to the
internal structure of dark matter (DM) halos. The~extent of this
minimum is influenced by the particle density threshold $\rho_0$. Notably,
DM halos exhibit nearly uniform density profiles, which can be
characterized by both the Navarro--Frenk--White (NFW) profile
\citep{Navarro:1997dq} and the
Einasto profile \citep{Einasto:2024aa}:
\begin{equation}
\rho(a) = \rho_0 \exp\left(-(a/a_c)^{1/N}\right).
\label{explaw}
\end{equation}
In  this context, $\rho_0$ represents the central density, while $a$
denotes the semi-major axis of the equidensity
ellipsoid. The~characteristic radius is indicated by $a_c$, and~$N$
serves as a 
structural parameter that allows for variations in the density
profile's shape. Research by \citet{Wang:2020ab} demonstrated that the
density profiles of halos across a diverse range of masses exhibit a
consistent shape parameter value of $\alpha = 1/N = 0.16$, maintaining
a similar form throughout a broad spectrum of halo masses. At~the
outer boundary of the halo, the~gradient transitions to $d \log \rho /
d \log r = -3.0$. It is important to highlight that the depth of the
minimum in the $\Lambda$CDM model sample aligns with the local value
of the $\gamma(r)$ function; thus, interpreting this as the fractal
dimension $D(r)$ may not be entirely~accurate.

Following the minimum observed at higher separation values $r$,
the~distribution of dark matter (DM) particles within filaments
outside 
the halos becomes predominant. This shift contributes to an increase
in the fractal dimension function. As~illustrated in
Figure~\ref{fig:corrFig3} and Figure~2 
 of \citet{Zehavi:2004aa}, the~transition from individual DM halos to the broader cosmic web occurs
at approximately  $r \approx 2$~\Mpc, which aligns well with the typical scales of
DM halos. Thus, we can conclude that the correlation functions of $\Lambda$CDM 
models uniquely characterize the internal structure of DM halos,
as~well as the fractal dimensional properties of the entire
cosmic~web. 

Figure~\ref{fig:corrFig3} illustrates that the fractal dimension
functions of SDSS galaxy samples closely resemble those of the
$\Lambda$CDM sample, albeit with significantly greater scatter
observed at small separations. The~minor discrepancies in shape
indicate that the internal structures of dark matter (DM) halos in the
$\Lambda$CDM models are distinct from those found in actual and
simulated galaxy clusters. In~the $\Lambda$CDM model samples, all DM
particles with density values $\rho \ge \rho_0$ are included, allowing
for a comprehensive view of the density profile of the halos extending
to their outer edges. Conversely, in~real galaxy samples, only
galaxies that exceed the selection threshold in brightness are
represented. Consequently, in~the most luminous galaxy samples, it is
common for only one or a few of the brightest galaxies to fall within
the observable range, leaving the true internal structures of the
clusters, up~to their outer boundaries, obscured.

\section{Comparing Angular and Spatial Distributions of
  Galaxies  \label{2d3d}} 

Redshifts are known to be influenced by the local movements of
galaxies within clusters, a~phenomenon referred to as the
Finger-of-God (FoG) effect. Additionally, galaxies and clusters tend
to move towards gravitational attractors, as~described by the Kaiser
effect~\citep{Kaiser:1987aa}. To~mitigate the impact of the Kaiser
effect when computing correlation functions, \citet{Davis:1983ly}
recommended utilizing galaxy position and velocity data independently,
as detailed in Section~\ref{angular}. The~inversion described in
Equation~(\ref{xi}) presupposes that the spatial three-dimensional (3D) and
projected two-dimensional (2D) density fields exhibit statistical
similarity. This premise has been widely accepted within the
astronomical community, and~the application of Equations~(\ref{wp}) and~(\ref{xi}) for the calculation of 3D correlation functions has become
a standard~practice.

A visual assessment of the 2D and 3D density fields, illustrated in
\mbox{Figures~\ref{fig:Lick} and \ref{SDSSslice},} reveals significant
distinctions between the two. The~3D density field is primarily
characterized by the filamentary structure of the cosmic web, while
the 2D field exhibits a more random distribution. Consequently, the~accuracy of the standard method for calculating correlation functions
(CFs) in this context remains uncertain. To~establish the relationship
between the 2D and 3D CFs, it is essential to compare these functions
using the same dataset. The~findings from this comparison have been
documented by \citet{Einasto:2021ti} and are presented in this~work.

\subsection{Relation Between 2D and 3D Correlation~Functions}

To compare 2D and 3D correlation functions, \citet{Einasto:2021ti}  constructed the 2D density
fields on a $2048^2$ grid by integrating the 3D field, 
\begin{equation}
  \delta_2(x,y) = \int_{z_1}^{z_2}  \delta(x,y,z) \dd{z}.
  \label{2dens}
\end{equation}
In the next phase of their research, the~authors segmented the cubic
sample into $n$ sequentially arranged 2D sheets, each measuring
$L_0 \times L_0 \times L$~\Mpc.  Here, $L = L_0/n$ represents the
thickness of each individual sheet, while $n$ takes values of 1, 2, 4,
and so forth, up~to 2048, indicating the total number of sheets
created. For~each value of $n$, the~authors computed the 2D
correlation functions (CFs) for all $n$ sheets and subsequently
determined the average CF corresponding to each $n$.The sheet
corresponding to $n=1$ encompasses the entire sample along the
$z$-direction, with~a thickness of $L=L_0=512$~\Mpc. For~$n=2$,
the~thickness reduces to $512/2=256$~\Mpc, while for $n=2048$,
the~thickness is $L=512/2048=0.25$~\Mpc. Through this methodology,
the~authors were able to calculate 2D correlation functions across a
variety of particle density thresholds $\rho_0$ and 2D sample
thickness $L$ within the context of the $\Lambda$CDM model, as~well as
for different magnitude limits derived from the Millennium sample
referenced by \citet{Springel:2005}.

Two-dimensional correlation functions (CFs) are influenced by two key
parameters: the thickness of the sheets, defined as $L=512/n$~\Mpc,
and~the particle density threshold for $\Lambda$CDM samples, denoted
as $\rho_0$, or~the 
magnitude limit $M_r$  for Millennium samples. The~analysis reveals that
the 2D CFs exhibit a luminosity dependence that closely mirrors that
of 3D CFs. This indicates that the correlation functions of 2D samples
retain the luminosity dependence characteristic of their 3D
counterparts. As~luminosity increases, the~amplitude of the
correlation functions also rises, illustrating the well-established
biasing effect described by \citet{Kaiser:1984}.

\begin{figure}[H]
   
\resizebox{0.48\textwidth}{!}{\includegraphics*{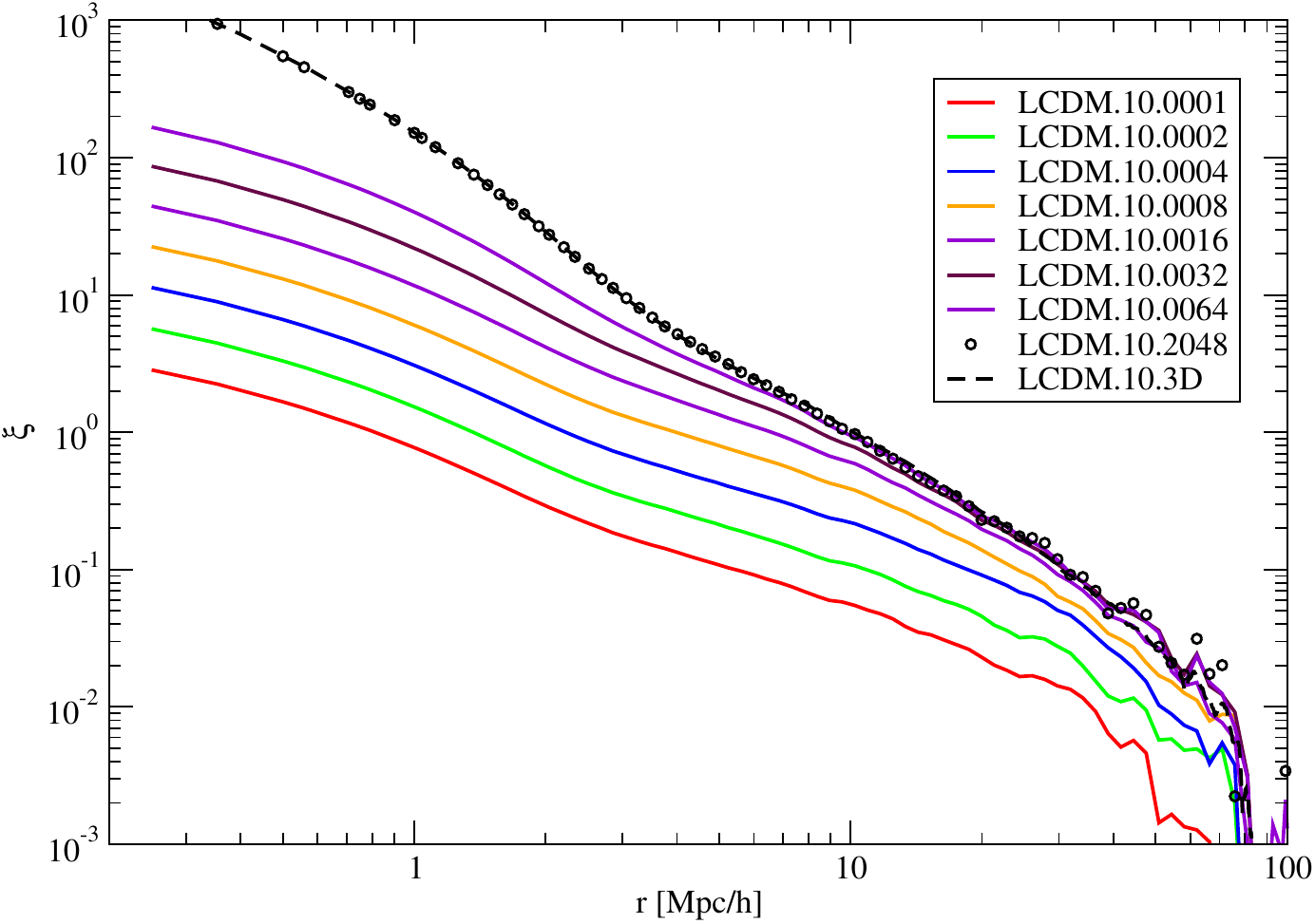}}
\resizebox{0.48\textwidth}{!}{\includegraphics*{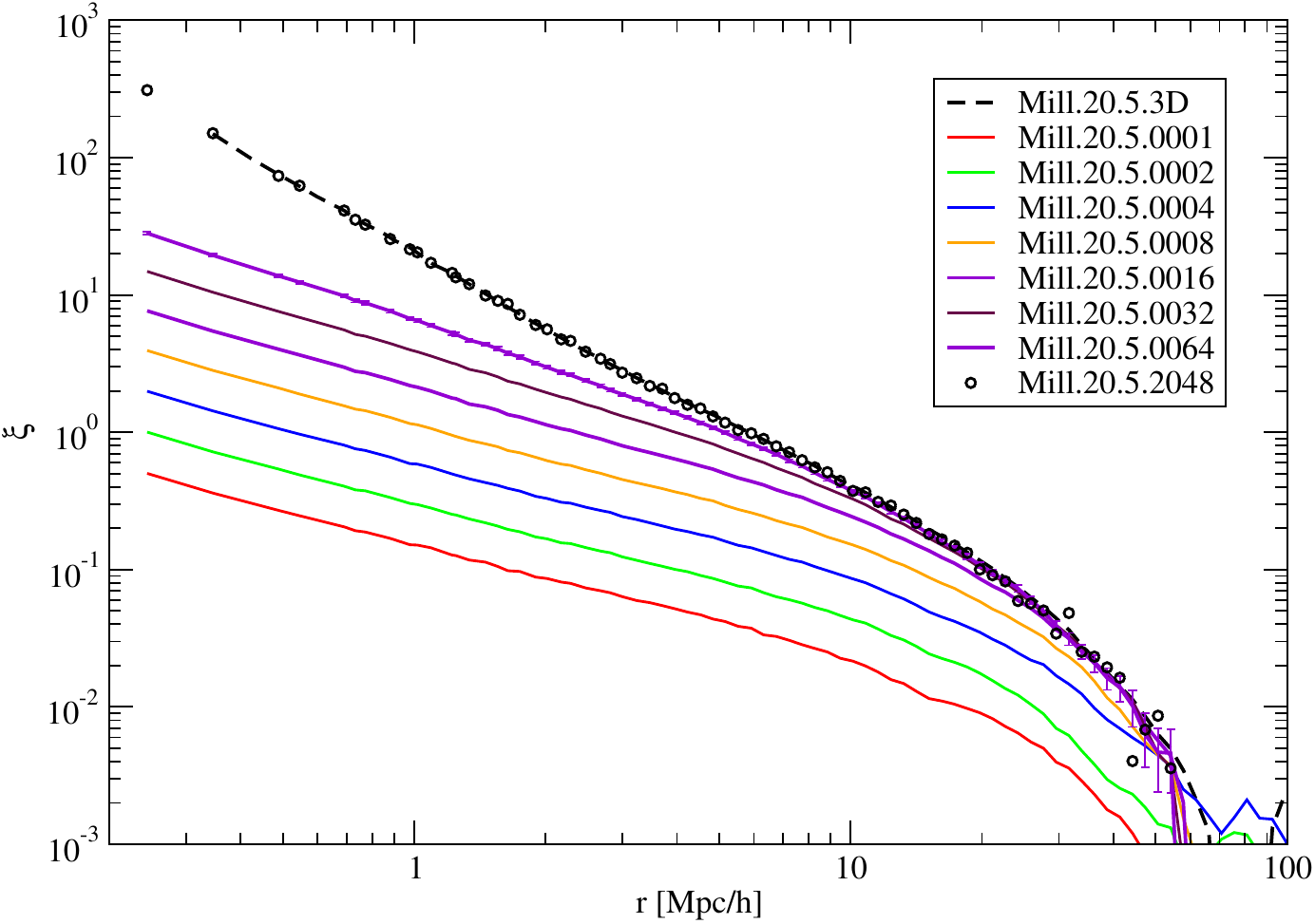}}
\caption{(\textbf{Left}):
 The two-dimensional correlation functions (CFs) of the
  $\Lambda$CDM model are presented with a particle density threshold
  of $\rho_0=10$, analyzed across various thicknesses of 2D
  samples. (\textbf{Right}): The two-dimensional CFs of the Millennium samples,
  constrained by a magnitude limit of $M_r=-20.5$. Different
  thicknesses $L$ of the 2D samples are denoted by lines of distinct
  colors. Dotted lines represent the
  three-dimensional CFss for samples with the same density
  threshold of $\rho_0=10$ and magnitude limit of $M_r=-20.5$
  \citep{Einasto:2021ti}. }
\label{fig:Fig6A} 
\end{figure}

In our study, a~key aspect is the relationship between the thickness
of the samples, $L$,  and the amplitude of 2D correlation functions.
Figure~\ref{fig:Fig6A} illustrates these correlation functions for a fixed
particle density limit of $\rho_0=10$ in LCDM.10 samples, as~well as
for Millennium samples Mill.20.5 with a luminosity threshold of
$M_r=-20.5$. These limits roughly align with $L^\ast$
galaxies. In~Figure~\ref{fig:Fig6A}, we present the 2D correlation
functions across 
various sample thicknesses, represented as $L=L_0/n$~\Mpc, with~the
number of sheets ranging from $n=1$ to $n=2048$. The~case with $n=1$
reflects the total sample thickness of $L=L_0$ and exhibits the lowest
amplitude. Conversely, the~final case corresponds to the average 2D
correlation function of the thinnest sheets, each measuring
$L=0.25$~\Mpc. Notably, the~2D correlation functions for the thinnest
samples, where $n=2048$, closely resemble the 3D correlation functions
indicated by the dotted lines in Figure~\ref{fig:Fig6A}. This finding
suggests that the structural information regarding dark matter halos
and the overall cosmic web is comprehensively retained in the {\em thin} 2D
correlation~functions.

\textls[-25]{The relationship between luminosity and correlation
  functions, as~discussed in this and the preceding section, is well
  established. Our 
analysis reveals that the amplitudes of two-dimensional correlation
functions (2D CFs) are affected by an additional parameter: the
sample thickness, denoted as $L$. The~variation in the amplitudes of 2D
CFs with respect to sample thickness is a consequence of the spatial
configuration of the cosmic web. This cosmic web comprises galaxies
arranged in a complex filamentary structure, leaving significant
regions of space unoccupied by galaxies. In~projected views, clusters
and filaments occupy these voids, which varies with sample
thickness. Consequently, the~patterns of the cosmic web in two
dimensions differ qualitatively from those in three dimensions,
with~the disparity becoming more pronounced as the thickness of the 2D 
sheets~increases.}

\subsection{Fractal  Analysis of the 2D Cosmic~Web \label{fractalweb}}

\begin{figure}[H]
\resizebox{0.48\textwidth}{!}{\includegraphics*{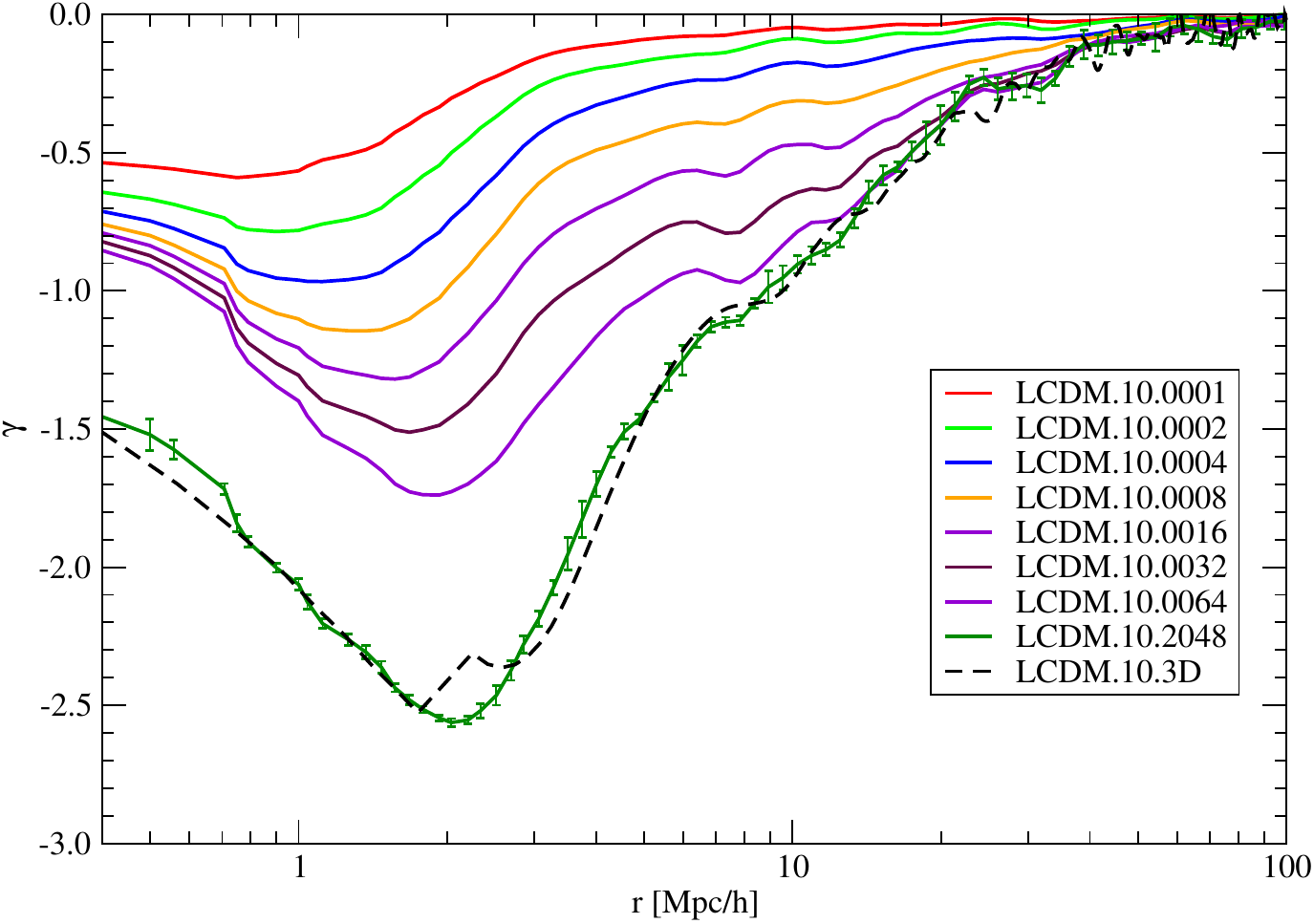}}
\resizebox{0.48\textwidth}{!}{\includegraphics*{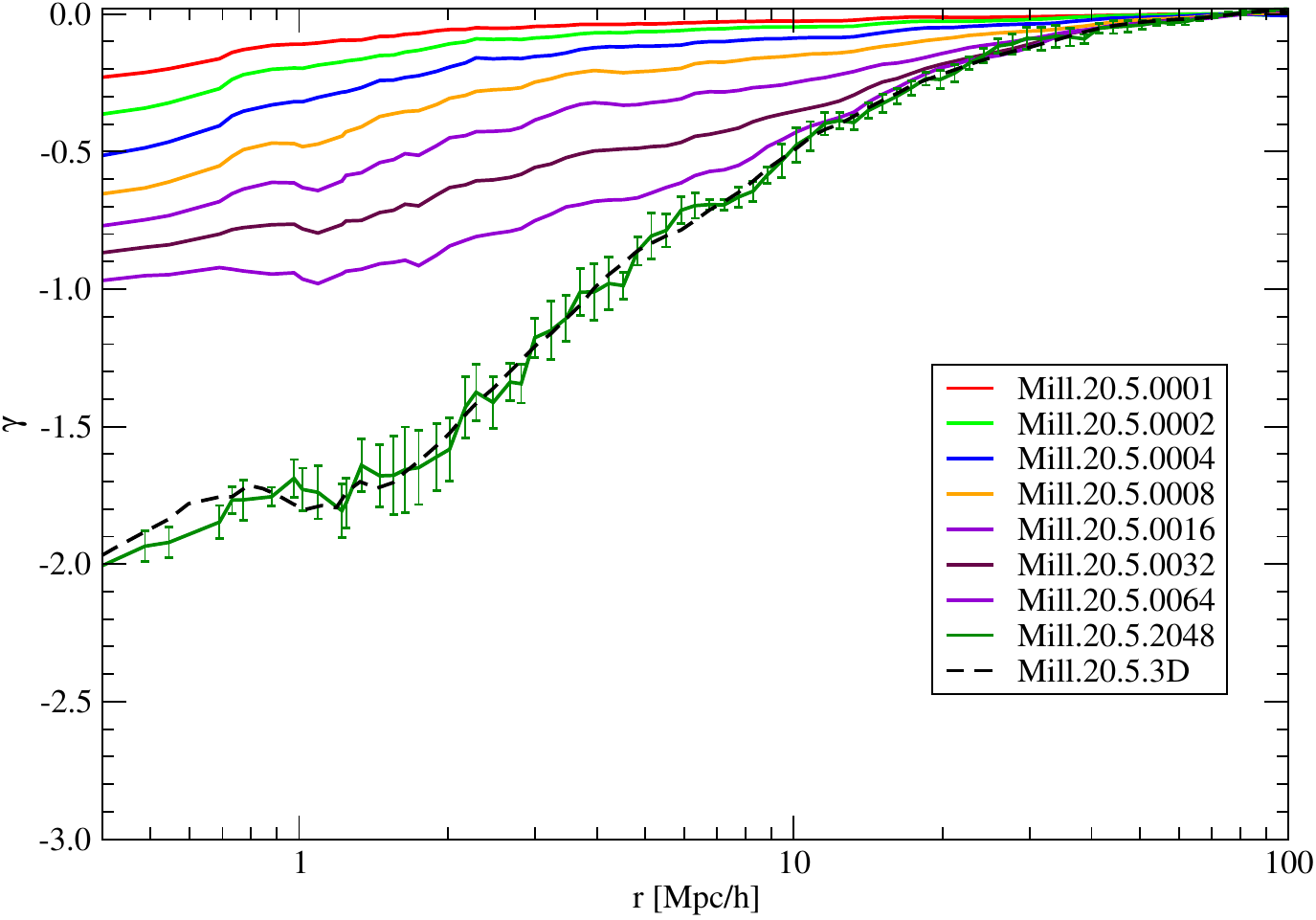}}
\caption{(\textbf{Left}):
 Two-dimensional gradient functions of the $\Lambda$CDM
  model, utilizing a particle density threshold of $\rho_0=10$, are
  presented for various thicknesses of 2D samples. (\textbf{Right}):
  Two-dimensional gradient functions of Millennium samples,
  constrained by a magnitude limit of $M_r=-20.5$, are displayed in
  real space
    \citep{Einasto:2021ti}. }
\label{fig:Fig6B} 
\end{figure}

In Figure~\ref{fig:Fig6B}, we illustrate the gradient functions for
the $\Lambda$CDM model, utilizing a particle density limit of
$\rho_0=10$, alongside the Millennium samples constrained by a
magnitude limit of $M_r=-20.5$. The~analysis employs pair separations
that are perpendicular to the line of sight, defined as
$r_p= \sqrt{(\Delta\,x)^2+(\Delta\,y)^2}$. The~parameter representing the thickness of the
samples, denoted as $L$, is also incorporated. Different thicknesses of
the 2D samples are indicated by lines of varying colors. For~comparative purposes, we include 3D functions represented by dotted
lines for samples with the same density and magnitude limits of
$\rho_0=10$ and $M_r=-20.5$, respectively. Additionally, error bars
are provided for the 2D samples where $n=2048$.

The comparison of the gradient functions of two-dimensional samples
with the fractal dimension functions of three-dimensional samples,
as~illustrated in Figure~\ref{fig:corrFig3}, indicates that the fine
structure information at small scales is retained only in the
two-dimensional samples derived from the $\Lambda$CDM
model. Conversely, in~the Millennium samples, the~details regarding
the internal structure of clusters are diminished in the
two-dimensional correlation functions (CFs). At~the luminosity
threshold of $M_r = -20.5$, the~clusters are comprised of only a
limited number of bright galaxies. The~amplitudes of the
two-dimensional CFs in the Millennium samples are relatively low,
leading to the dominance of the first constant term in the gradient
function $g(r) = 1+\xi(r)$.  As~depicted in the right panel of
Figure~\ref{fig:Fig6B}, for~thicker samples, the~slope of the
two-dimensional CF exhibits a gradual variation over an extensive
range of separations, specifically for $r_p > 1$~\Mpc. This behavior
supports the established observation that the two-dimensional CF can
be effectively modeled using a simple power-law function,
as~demonstrated by \mbox{\citet{Groth:1977aa},}
\mbox{\citet{Davis:1983ly},} and \citet{Maddox:1990aa}. 
Figure~\ref{fig:Fig6B} illustrates that the value of the
two-dimensional gradient function depends on  the thickness of
the samples. When the thickness ranges from $64$ to $128$~\Mpc,
the~gradient achieves a value of approximately $\gamma(r) \sim -0.7$
at shorter distances, smoothly approaching $\gamma(r) = 0$ at
$r = 100$~\Mpc. This finding explains  the results reported by
\citet{Groth:1977aa}, \citet{Davis:1983ly}, and \citet{Maddox:1990aa}
concerning the angular correlation~function.

\textls[-15]{The previous examination of the correlation study
  conducted by \mbox{\citet{Groth:1977aa}} and \citet{Peebles:2001aa}
  revealed a flat profile of 2D correlation functions across an
  extensive separation range of $0.05 \le r \le 9$~\Mpc. Our analysis
  suggests that the 2D CFs are not sensitive to the presence of halos
  (clusters), rendering the actual structure at smaller separations
  undetectable. Additionally, the amplitude of the 2D correlation
  functions plays a significant role. As~illustrated in
  Figure~\ref{fig:Fig6A}, the~amplitude of the 2D CFs is always lower
  than that of the 3D CFs, which varies with the thickness of the 2D
  samples. According to \mbox{\citet{Norberg:2001aa}} and
  \mbox{\citet{Zehavi:2005aa}}, the~correlation lengths for the
  faintest galaxies were measured at approximately
  $r_0\approx 4.5$~\Mpc, a~value interpreted as representative of 3D
  samples. Figure~\ref{fig:Fig6A} suggests that for these faint
  galaxies, the~actual amplitude of the 3D correlation functions is
  greater, resulting in true 3D correlation lengths near 6~\Mpc. This
  finding aligns closely with our measurements of the correlation
  lengths of SDSS samples, as~depicted in Figure~\ref{fig:corrFig6}.}

\section{Structure and Evolution of Cosmic Web from Combined Spatial 
  and Velocity Data \label{velocities}} 

The correlation function used in the study of fractal properties of
the cosmic web uses only spatial data on the distribution of galaxies
and dark matter. Modern numerical simulations and observational data
allow the use of all phase--space data---spatial positions and velocities of
particles and galaxies. In~this section, I  discuss the structure and
evolution of the cosmic web using full phase--space data. Such combined data
are very useful to study the hierarchy of the cosmic web
in low-density regions---voids.


\subsection{Void~Hierarchy}

One aspect of the hierarchical structure of the cosmic web is the
hierarchy of voids.  Already early studies showed that diameters of
voids have a large scatter, from~a few megaparsecs to hundred
megaparsecs (\citet{Kirshner:1981}, \citet{Pan:2012aa},
\citet{Sutter:2012aa}, \citet{Nadathur:2014aa}). \citet{Sheth:2004ly}
studied the formation and evolution of voids, using numerical
simulations of the evolution of the cosmic web. The authors showed that
voids have a remarkable hierarchical structure---voids are filled
with a complex web of tenuous filaments and low-mass
haloes. During~the evolution, larger voids grow by the mergers of
smaller voids, which 
is analogous to how massive clusters form by merging less massive
clusters and groups. Small voids, which are located on overdensity
regions, disappear as the overdensity collapses around~them.

\begin{figure}[H]
\resizebox{0.85\textwidth}{!}{\includegraphics*{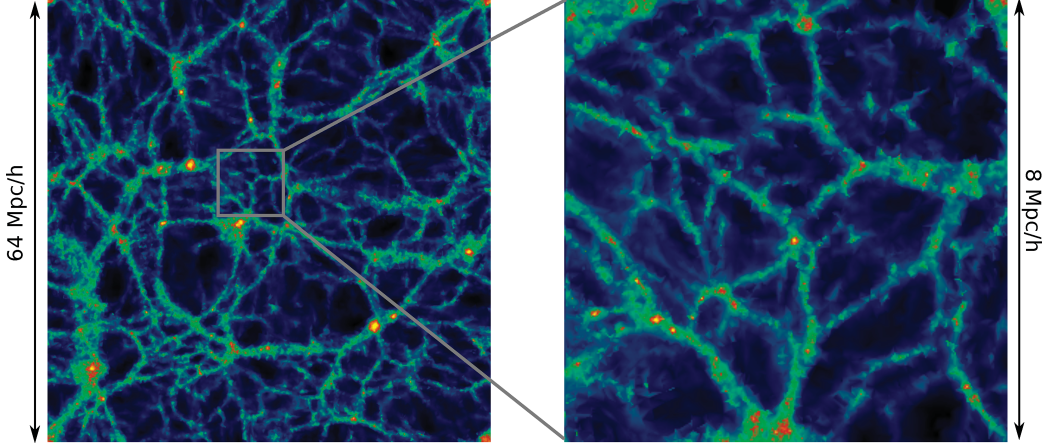}}
\caption{The 
 hierarchy of structure within the cosmic web is
  illustrated in the provided visuals. The~left panel displays the
  density field across a narrow slice of a 64~\Mpc\  simulation based on
  the $\Lambda$CDM cosmology model. Meanwhile, the~right panel focuses
  on a specific highlighted area within that slice. The~density field
  for this zoomed-in region was derived from a high-resolution
  resimulation featuring $N_{part} = 1024^3$, centered in the void
  region  \citep{Aragon-Calvo:2013aa}.}
\label{AragonFig1}
\end{figure}

In their study, \citet{Aragon-Calvo:2013aa} conducted a thorough
examination of the hierarchical organization of cosmic voids through
advanced numerical simulations. The~large-scale fluctuations
responsible for the formation of voids encompass smaller fluctuations
that develop within regions resembling a locally low-density
Universe. This phenomenon is evident at every level within the void
hierarchy, as~subvoids themselves harbor even smaller sub-subvoids,
as~illustrated in Figure~\ref{AragonFig1}. This observation highlights
the 
hierarchical nature of the cosmic web, where low-density filamentary
structures on smaller scales exhibit similarities to those on larger
scales.

\begin{figure}[H]
\resizebox{0.70\textwidth}{!}{\includegraphics*{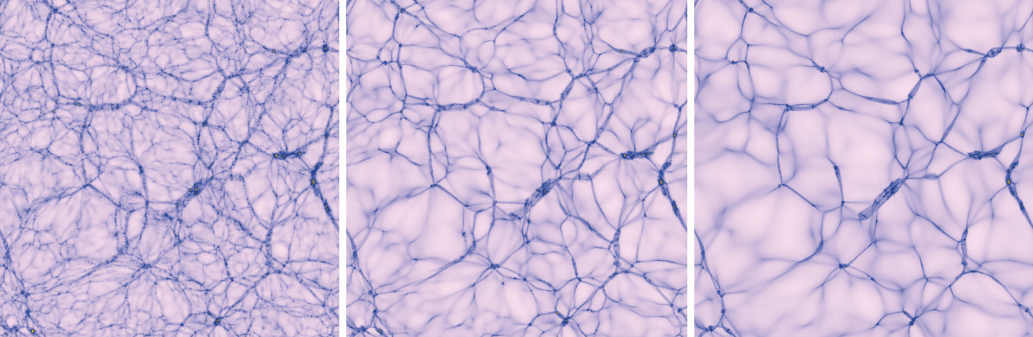}}\\
\resizebox{0.70\textwidth}{!}{\includegraphics*{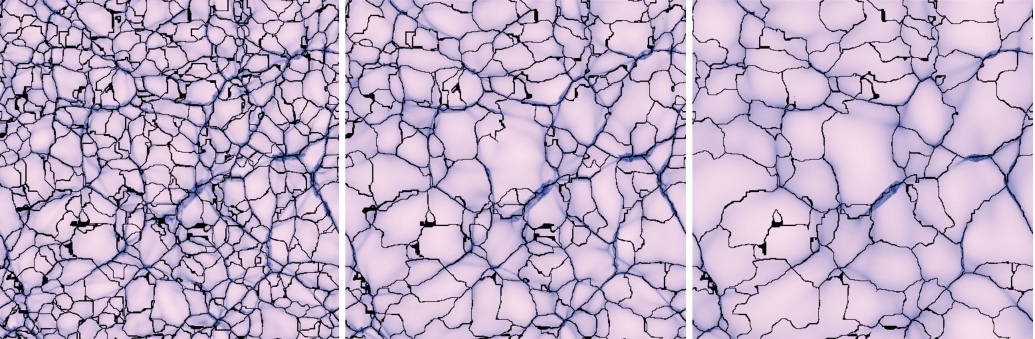}}
\caption{{\textbf{Top}):} 
 Density field across a thin slice of the simulation box
  at $z =0$ for three cases: original field (\textbf{left} panel), after~smoothing with 2~\Mpc\ (\textbf{central} panel), and with 4~\Mpc\ (\textbf{right}
  panel). (\textbf{Bottom}): Hierarchical cosmic web spine superimposed to its
  corresponding density field for the bottom, middle, and top levels
  (left, central, and right panels, respectively)
  \citep{Aragon-Calvo:2013aa}. }

\label{AragonFig2}
\end{figure}
\unskip

\begin{figure}[H]

\resizebox{0.65\textwidth}{!}{\includegraphics*{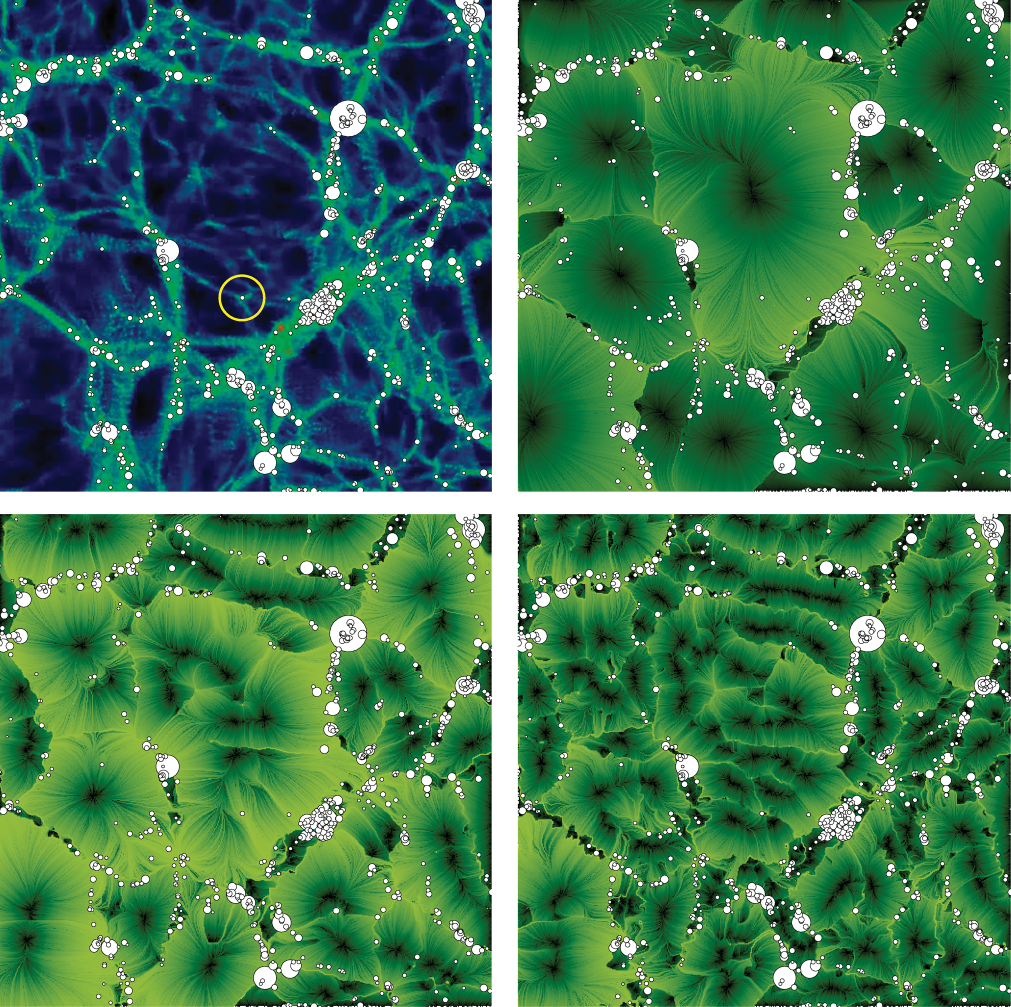}}
\caption{Dark 
 matter haloes superimposed on the density and velocity
  fields. \mbox{The~(\textbf{top-left} panel)} shows a slice of the density field and 
   all the FoF haloes closer than 1~\Mpc\ from the
  slice. The~residual velocity field is shown at scales 4, 2, and 1
  \Mpc\ in the (\textbf{top-right}), (\textbf{bottom-left}), and (\textbf{bottom-right} panel),
  respectively. We show the same FoF haloes as in the density field
  slice~\citep{Aragon-Calvo:2013aa}.}
\label{AragonFig3}
\end{figure}
\unskip

Cosmic structures of varying scales can be analyzed by applying
smoothing techniques to density fields with different smoothing
parameters. The~original density field retains all intricate details,
while smoothing at a scale of 2~\Mpc\ emphasizes structures
characteristic of galaxy groups. In~contrast, a~smoothing scale of
4~\Mpc\ brings out features such as the cores of superclusters and
large voids. The~top panels of Figure~\ref{AragonFig2} illustrate the
original density field at redshift $z=0$ (left panel), alongside its
smoothed versions at 2 and 4~\Mpc\ (center and right panels,
respectively). By~comparing the panels with different smoothing
scales, we can observe the hierarchical arrangement of filaments and
voids. To~capture the finer details of this cosmic web, the~boundaries
of voids were identified using the SpineWeb method, which is depicted
in the bottom panels of Figure~\ref{AragonFig2}. The~SpineWeb method,
as~described by \citet{Aragon-Calvo:2013aa}, involves calculating the 
density and velocity fields, along with their respective~gradients.

It is widely recognized that galaxies exclusively develop within dark
matter halos, as~void regions lack sufficient density for galaxy
formation. Figure~\ref{AragonFig3} illustrates the variations in the
distribution of simulated galaxies (dark matter halos) alongside the
density and velocity fields. The~top-left panel displays a slice of
the density field, highlighting halos within a 2~\Mpc\ thick
section. These halos are situated in high-density areas of the cosmic
web, including filaments and clusters. The~process of galaxy formation
occurs in two stages: initially, dark matter aggregates to form halos,
followed by the emergence of galaxies within these halos
\citep{White:1978}. As~depicted in Figure~\ref{AragonFig3}, halos
containing galaxies occupy only a minor portion of the overall spatial
volume.

The velocity fields depicted in the top right and bottom panels of
Figure~\ref{AragonFig3} are presented at smoothing scales of 4, 2,
and~1~\Mpc, alongside the same halos illustrated in the top left 
panel. This figure illustrates the hierarchy of voids nested within
larger voids, where the largest voids are subdivided into
progressively smaller ones. The~local velocity fields in regions
surrounding halos and voids exhibit significant differences. In~halo
regions, the~velocity field is characterized by turbulence, which
facilitates the condensation of matter into halos and
subhalos. Conversely, at~the larger smoothing scales of 4 and 2~\Mpc,
the velocity field surrounding the void halo is predominantly laminar
and directed toward the halos. At~the smaller scale of 1~\Mpc,
the~velocity field approaches the halo from multiple directions. It is 
important to note that this intricate structure observed in
low-density regions is composed of a sparse field of dark matter
filaments and baryonic gaseous~matter.

\subsection{Evolution of Galaxies in the Void~Hierarchy}

Most galaxy formation models successfully replicate a diverse array of
observations and shed light on the physical processes taking place
within halos as distinct entities~\citep{White:1978}. In~reality,
galaxies develop within the large-scale environment of the cosmic web. To~comprehend how the cosmic environment influences galaxy
characteristics, \mbox{\citet{Aragon-Calvo:2019aa}} introduced the Cosmic Web
Detachment (CWD) model. This model integrates multiple mechanisms that
inhibit star formation and illustrates how galaxies acquire
star-forming gas in their formative stages through a network of
primordial filaments, as~outlined in the preceding~section.

\begin{figure}[H]

\resizebox{0.80\textwidth}{!}{\includegraphics*{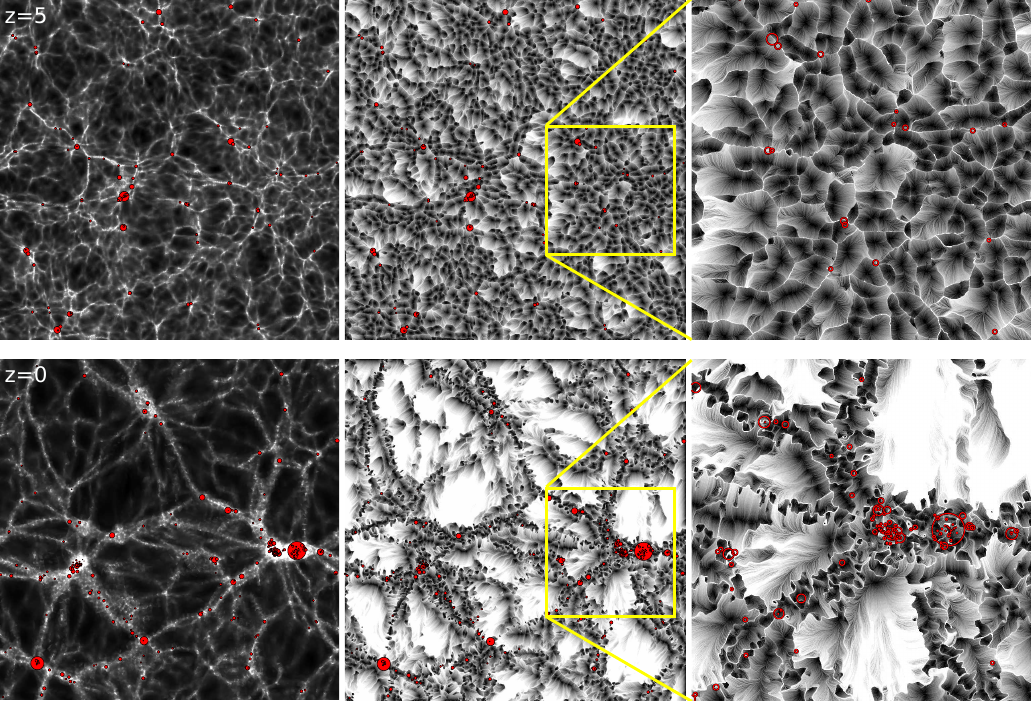}}
\caption{Coherent 
 vs. chaotic velocity field around halos. For~explanation see text \citep{Aragon-Calvo:2019aa}.}
\label{AragonFig1A}
\end{figure}

The development and progression of hierarchical structures within the
cosmic web are illustrated in Figure~\ref{AragonFig1A}. The~upper panels
depict the density and velocity fields of the cosmic web at an early
epoch, specifically at redshift $z=5$, while the lower panels
represent the current epoch at $z=0$. The~left panels showcase density
fields spanning 32~\Mpc, the~central panels display the corresponding
velocity fields, and~the right panels provide a cross section of these
fields. During~the early epoch, the~velocity field exhibits coherence,
with star-forming cold gas being accreted through primordial  coherent
filamentary streams. Star formation occurs in conjunction with the
accretion of cold gas and ceases once the gas supply is depleted. In~contrast, at~later epochs, the~velocity field surrounding
halos—indicated by red and white circles—becomes highly~chaotic.

\unskip

\section{Scale of~Homogeneity \label{scale}}

A key question in the fractal analysis of the cosmic web centers
around the scale of homogeneity, where opinions among various authors
diverge significantly. The~Anglo-American school, as~discussed in
Section~\ref{angular}, has predominantly focused on utilizing only 2D
data for their studies. According to 
\citet{Maddox:1990aa}, the~fractal nature of galaxy distribution
appears to be applicable within the distance range of 10 kpc $\le r
\le 10$~\Mpc, exhibiting a fractal dimension of approximately $D
\approx 1.3$. This upper limit has been regarded as the scale of
homogeneity.

The Italian School, as~noted by \citet{Pietronero:1987aa} and
\citet{Sylos-Labini:1996aa}, established that the fractal nature of
galaxy distribution, characterized by a fractal dimension of
approximately $D \approx 1.7$, holds true from small scales up to the
most extensive scales examined for visible matter. The samples studied
by \citet{Sylos-Labini:1996aa} encompass 
radio galaxies and quasars, spanning a magnitude range of $12 \le m
\le 28$.

One method for determining the limit of a fractal structure involves
calculating the fractal dimension function from the gradient function,
as outlined in Section~\ref{web}. The~$\Lambda$CDM and SDSS fractal
dimension functions, illustrated in Figure~\ref{fig:corrFig3}, converge
towards the limit of $D=3$, at a distance of approximately $r \approx
100$~\Mpc, which represents just 1/5 of the sample
size. \citet{Pan:2000aa} examined the spatial distribution of IRAS
sources from the PSC catalog and derived fractal dimensions across two
distance ranges: \mbox{$20 < r < 50$}~\Mpc\ and $r > 50$~\Mpc. In~the first
range, the~fractal dimension fluctuated between $2.05 \le D \le 2.83$,
while in the second range, it remained constant at
$D=3.0$. \citet{Sarkar:2009aa} applied multifractal analysis to assess
the scale of homogeneity within the range of 60 to 70~\Mpc, utilizing
SDSS DR6 spectroscopic galaxy data. Furthermore,
\citet{Scrimgeour:2012aa} conducted a spectroscopic survey of blue
galaxies within a cosmic volume of approximately 1\Gpc\ and established
a lower limit for the fractal dimension of $D_2=2.97$ on scales
ranging from about 80~\Mpc\ to $\sim$300~\Mpc. 

One approach to determining the scale of homogeneity is to examine the
structures of the largest astronomical objects, such as voids and
superclusters. The~recent catalog of superclusters from the SDSS
survey compiled by \citet{Liivamagi:2012} includes objects identified
using both adaptive and fixed density thresholds. The~largest
superclusters reach sizes of up to 120~\Mpc\ for the primary galaxy
superclusters and 200~\Mpc\ for the LRG superclusters. Notably,
the~largest superclusters recorded in the \citet{Sankhyayan:2023aa} 
catalog of SDSS superclusters also measure 200~\Mpc.

Independent insights into the structure of the cosmic web are derived
from velocity data. In~their analysis, \citet{Courtois:2025aa}
examined the velocity field utilizing cosmic flow CF4 peculiar
velocities. The~research revealed that the bulk flow amplitude
approaches zero at greater distances, suggesting an increasing
homogeneity of the Universe. Within~a range of 150~\Mpc, the~measured
bulk flow is $230\pm 136$ km/s. This indicates that the dynamic scale
of homogeneity has not yet been attained within the 200--300~\Mpc\ 
interval from the observer, signifying that the local Universe
continues to display notable fluctuations in mass distribution and the
dynamics of galaxy movements.  Additionally, \citet{Park:2015aa}
investigated a sample of quasars from the SDSS survey, discovering
that this sample deviates from a random Poisson
distribution. The~authors concluded that the concept of a scale of
homogeneity is achieved only asymptotically as
the observational scale increases, with~no inherent characteristic
scale in the Universe beyond which it can be considered
homogeneous. Nevertheless, at~scales exceeding 300~\Mpc,
the~distribution of quasars approaches that of a homogeneous~sample.

\section{Summary and~Outlook \label{summ}}

The cosmic web is a complex geometric pattern.  One of the aspects of
the structure of the cosmic web is its fractal nature, which was
recognized already in the introduction of the fractal concept by
\citet{Mandelbrot:1977aa, Mandelbrot:1982uq}.  In~this review, I 
discussed various aspects of fractal properties of the cosmic web from an
observational point of view.   Our discussion  can be summarized  in
following~points.

{\bf {Fractal properties from two-dimensional data}}.  The~first
application of the fractal character of the distribution of galaxies
was made by \citet{Soneira:1978fk} in the construction of the angular
distribution of galaxies in a fractal way to mimic the
\citet{Shane:1967} distribution of galaxies, as~displayed in
Figure~\ref{fig:Lick}.  A~deeper 2D distribution of APM galaxies was
expressed by \citet{Maddox:1990aa} by a power-law correlation
function, which has a constant slope $-1.7$ over the range of angular
distances, $0.01 \le \theta \le 3$~degrees.  This was interpreted as a
hint that the power-law correlation function is valid in the range of
separations 10 kpc$ \le r \le 10$~\Mpc\ \citep{Peebles:1989aa}.

{\bf {Determining fractal dimension.}}  As discussed in
Section~\ref{dimens}, \citet{Pietronero:1987aa}  noticed that the correlation
function is normalized to a Poissonian distribution and~
is forced to vanish at large scales. For~this reason, it is not suited
for measuring large-scale homogeneity.  To~measure the fractal
dimension, instead of the correlation function, its derivative, the~structure function, $g(r)=1+\xi(r)$, and~its log--log gradient,
$\gamma(r)= {d \log g(r) \over d \log r}$, should be used.  The~fractal dimension can be found from the gradient as follows: $D(r) =3+
\gamma(r)$. In~our analysis, we have used this definition of the
fractal~dimension.

{\bf {Fractal properties from three-dimensional data.}}  Essential
fractal properties of the cosmic web are displayed in the fractal
dimension function,  Figure~\ref{fig:corrFig3}.  The~analysis was based
on a $\Lambda$CDM model of size 512~\Mpc, and~a SDSS sample of similar
volume.  The~fractal dimension function of both samples has two well
separated regions: on~small separations,  $r\le 3$~\Mpc,  the~function
characterizes the distribution of particles/galaxies in halos, and on~larger separations, it characterizes the distribution of particles/galaxies in
filaments.  Fractal dimension functions depend on the magnitude
(particle density) limits of samples. The~ minimum of
the dimension function at scale $r\approx 2$~\Mpc\ is deeper for
samples of luminous galaxies. However, the~depth of the minimum of the
fractal dimension function is exaggerated, since it is based on the
local value of the gradient of the structure function, $g(r)=1+\xi(r)$. 

The gradient function from 2D data, presented in Figure~\ref{fig:Fig6B},
depends on the depth of the 2D sample.  Very thin 2D samples behave
similar to 3D samples. With~increasing thickness of the samples, the
information on the distribution of particles/galaxies in halos is
gradually erased. This effect is very strong for SDSS and Millennium
galaxy samples.  Analyses of the relation between 2D and 3D
correlation functions, presented in Section~\ref{2d3d}, shows that, in
2D distribution of galaxies, the information on the distribution of
galaxies in clusters has been erased, and~the division of the
correlation function into two regions, halos and filaments, is not
seen.

{\bf {Fractal properties from velocity data.}}  Velocity data yield
essential additional information on the structure of the cosmic
web. The~combination of spatial and velocity data shows that the
internal structure of voids is very complex.  Inside voids there
exist subvoids, sub-subvoids, etc., and~the fractal character of the
dark matter distribution continues to small~scales.

{\bf {Scale of homogeneity.}}  Early 2D data emphasized that the fractal
character of the distribution of galaxies extends only to
$\sim 10$~\Mpc\  and~that beyond this limit, the distribution of
galaxies is homogeneous.  Later analyses have shown that the
correlation function $\xi(r)$ is not suited to find the limit of the fractal
nature of the galaxy distribution; instead, the structure function
$g(r)=1+\xi(r)$ can be applied. The~scale of homogeneity has been
studied by many authors, who found that the local Universe still has
some fluctuations in the distribution of galaxies on distance
$\approx200$~\Mpc. Homogeneity is only achieved
asymptotically, as the scale of observation~increases.

To conclude, we can say that the contemporary understanding of the fractal
properties of the Universe includes the best aspects of both the
Anglo-American and Italian~approaches. 

\vspace{6pt}

\funding{This work was supported by  Tartu Observatory, University
  of~Tartu. }

\dataavailability{No new data were created or analyzed in this study. Data sharing is not applicable to this article.}

\acknowledgments{Our special thanks to the anonymous referees for
  stimulating suggestions that greatly improved the paper and~to
  colleagues in Tartu Observatory for discussions.  Figures  are
  reproduced by permission of the AAS, Monthly Notices of the Royal
  Astronomical Society, and~EDPsciences. Permission is granted for the purpose of reuse
  in “Fractal properties of the cosmic web” and~does
  not extend to any other forms of distribution or reproduction beyond
  what is customary for the journal's dissemination practices.
}

\conflictsofinterest{The author declares no conflicts of interest. }


%




\begin{adjustwidth}{-\extralength}{0cm}

\reftitle{References}

\PublishersNote{}
\end{adjustwidth}

%


\end{document}